\documentclass[11pt]{article}
\usepackage[dvips]{graphicx}
\usepackage{epsfig}
\usepackage{amssymb}
\usepackage{color}
\usepackage[left]{lineno}

\begin{document}
\centerline{\LARGE EUROPEAN ORGANIZATION FOR NUCLEAR RESEARCH}
%
%
\vspace{20mm} {\flushright{
CERN-PH-EP-2012-367 \\
17 December 2012
\\}}
\vspace{-35mm}
%
%
\vspace{35mm}

\begin{center}
{\bf {\Large \boldmath{Precision Measurement of the Ratio of the\\
Charged Kaon Leptonic Decay Rates}}}
\end{center}
\begin{center}
{\Large The NA62 collaboration$\,$\renewcommand{\thefootnote}{\fnsymbol{footnote}}%
\footnotemark[1]\renewcommand{\thefootnote}{\arabic{footnote}}}\\
\end{center}

\begin{abstract}
A precision measurement of the ratio $R_K$ of the rates of kaon
leptonic decays $K^\pm\to e^\pm\nu$ and $K^\pm\to\mu^\pm\nu$ with
the full data sample collected by the NA62 experiment at CERN in
2007--2008 is reported. The result, obtained by analysing $\sim\!
150000$ reconstructed $K^\pm\to e^\pm\nu$ candidates with 11\%
background contamination, is $R_K=(2.488\pm0.010)\times10^{-5}$, in
agreement with the Standard Model expectation.
\end{abstract}

\begin{center}
\it{Accepted for publication in Physics Letters B\\18 January 2013}
\end{center}

\setcounter{footnote}{0}
\renewcommand{\thefootnote}{\fnsymbol{footnote}}
\footnotetext[1]{Copyright CERN for the benefit of the NA62
collaboration. Contact: E. Goudzovski, eg@hep.ph.bham.ac.uk.}
\renewcommand{\thefootnote}{\arabic{footnote}}

\newpage
\begin{center}
{\Large The NA62 collaboration}\\
\vspace{2mm}
 C.~Lazzeroni$\,$\footnotemark[1],
 A.~Romano\\
{\em \small University of Birmingham, Edgbaston, Birmingham,
B15 2TT, United Kingdom} \\[0.2cm]
 A.~Ceccucci,
 H.~Danielsson,
 V.~Falaleev,
 L.~Gatignon,
 S.~Goy Lopez$\,$\footnotemark[2],
 B.~Hallgren$\,$\footnotemark[3],
 A.~Maier,
 A.~Peters,
 M.~Piccini$\,$\footnotemark[4],
 P.~Riedler\\
{\em \small CERN, CH-1211 Gen\`eve 23, Switzerland}
% ...
\\[0.2cm]
\end{center}
%
%
%
\footnotetext[1]{Supported by a Royal Society University Research Fellowship}
\footnotetext[2]{CIEMAT, E-28040 Madrid, Spain}
\footnotetext[3]{University of Birmingham, Edgbaston, Birmingham,
B15 2TT, United Kingdom}
\footnotetext[4]{Sezione dell'INFN di Perugia, I-06100 Perugia,
Italy}
\footnotetext[5]{Physikalisches Institut,
Ruprecht-Karls-Universit\"{a}t Heidelberg, D-69120 Heidelberg,
Germany}
\footnotetext[6]{Funded by the National Science Foundation under award No. 0338597}
\footnotetext[7]{CERN, CH-1211 Gen\`eve 23, Switzerland}
\footnotetext[8]{Also at Dipartimento di Fisica, Universit\`a di
Modena e Reggio Emilia, I-41125 Modena, Italy}
\footnotetext[9]{Also at Dipartimento di Fisica, Universit\`a di
Firenze, I-50019 Sesto Fiorentino (FI), Italy}
\footnotetext[10]{Also at Istituto di Fisica, Universit\`a di Urbino,
I-61029 Urbino, Italy}
\footnotetext[11]{Funded by the German Federal Minister for Education and Research (BMBF) under contract 05HA6UMA}
\footnotetext[12]{Dipartimento di Fisica dell'Universit\`a e Sezione
dell'INFN di Padova, I-35131 Padova, Italy}
\footnotetext[13]{Dipartimento di Fisica Sperimentale
dell'Universit\`a di Torino, I-10125 Torino, Italy}
\footnotetext[14]{Funded by Consejo Nacional de Ciencia y Tecnolog\'{\i}a {\nobreak (CONACyT)} and Fondo de Apoyo a la Investigaci\'on (UASLP)}
\footnotetext[15]{Funded by the Bulgarian National Science Fund
under contract DID02-22}
\footnotetext[16]{Universit\`a degli Studi del Piemonte Orientale,
I-13100 Vercelli, Italy}
\footnotetext[17]{University College Dublin School of Physics,
Belfield, Dublin 4, Ireland}
\newpage


\section*{Introduction}

The decays of pseudoscalar mesons to light leptons are helicity suppressed in the Standard Model (SM) due to the $V\!\!-\!\!A$ structure of the charged current coupling. In particular, the SM width of $P^\pm\to\ell^\pm\nu$ decays with $P=\pi,K,D,B$ (denoted as $P_{\ell 2}$ below) is
\begin{displaymath}
\Gamma^\mathrm{SM}(P^\pm\to\ell^\pm\nu) = \frac{G_F^2 M_P
M_\ell^2}{8\pi} \left(1-\frac{M_\ell^2}{M_P^2}\right)^2
f_P^2|V_{qq\prime}|^2,
\end{displaymath}
where $G_F$ is the Fermi constant, $M_P$ and $M_\ell$ are the meson and lepton masses, $f_P$ is the meson decay constant, and $V_{qq\prime}$ is the corresponding CKM matrix element. Although the SM predictions for $P_{\ell 2}$ decay rates are affected by hadronic uncertainties via the decay constant, ratios of decay rates of the same parent meson do not depend on $f_P$ and can be computed very precisely. In particular, the SM prediction for $R_K=\Gamma(K_{e2})/\Gamma(K_{\mu 2})$, inclusive of internal bremsstrahlung (IB) radiation, is~\cite{ci07}
\begin{displaymath}
R_K^\mathrm{SM} = \left(\frac{M_e}{M_\mu}\right)^2
\left(\frac{M_K^2-M_e^2}{M_K^2-M_\mu^2}\right)^2 (1 + \delta
R_{\mathrm{QED}})= (2.477 \pm 0.001)\times 10^{-5},
\end{displaymath}
where $\delta R_{\mathrm{QED}}=(-3.79\pm0.04)\%$ is the electromagnetic correction.

Within extensions of the SM involving two Higgs doublets, $R_K$ is sensitive to lepton flavour violating effects induced by loop processes with the charged Higgs boson ($H^\pm$) exchange~\cite{ma06}. A recent study~\cite{gi12} has concluded that $R_K$ can be enhanced by ${\cal O}(1\%)$ within the Minimal Supersymmetric Standard Model. However, the potential new physics effects are constrained by other observables such as $B_S\to\mu^+\mu^-$ and $B^+\to\tau^+\nu$ decay rates~\cite{fo12}. Moreover, $R_K$ is sensitive to the neutrino mixing parameters within SM extensions involving a fourth generation of quarks and leptons~\cite{la10} or sterile neutrinos~\cite{ab12}.

Measurements of $R_K$ have recently been reported by the KLOE~\cite{am09} and NA62~\cite{la11} experiments. An improved measurement based on the full dedicated data sample collected by the NA62 experiment in 2007--2008 and superseding the earlier result~\cite{la11} is reported here.

\section{Beam and detector}

\subsection{Beam line}
\label{sec:beam}

The beam line of the earlier NA48/2 experiment~\cite{ba07} was used for the NA62 data taking in 2007--2008. Either simultaneous or single beams of positive and negative secondary hadrons, with central momentum of 74 GeV/$c$ and momentum spread of $\pm1.4$~GeV/$c$ (rms), were derived from the primary 400 GeV/$c$ protons extracted from the CERN SPS and impinging on a 40~cm long, 0.2~cm diameter beryllium target. The beam momenta were selected by the first two magnets in a four dipole achromat and by momentum-defining slits incorporated into a 3.2~m thick copper/iron proton beam dump, which also provided the possibility of blocking either of the two beams. The beam composition was dominated by pions ($\pi^\pm$), with kaon ($K^\pm$) fractions of about 6\%. The $K^+$ and $K^-$ beams entered the decay fiducial volume at angles of $\pm0.23$~mrad ($\pm0.30$~mrad in the early stage of data taking, about 25\% of the total beam flux) with respect to the detector axis, so as to compensate for the opposite $\mp3.58$~mrad deflections by the downstream spectrometer magnet. These deflections were regularly reversed during the data taking. The individual beam particles were not tagged, and their momenta were not measured. The beam kaons decayed in a fiducial volume contained in a 114~m long cylindrical vacuum tank.

The hadron beams were accompanied by an intense flux of stray muons travelling outside the beam vacuum pipe. Two 5~m long magnetized iron toroids with small horizontal and vertical apertures centered on the beam line were installed upstream of the decay volume to suppress backgrounds associated with these ``halo'' muons. These toroids, named ``muon scrapers'', were operated with the same magnetic field polarity chosen to deflect positive halo muons away from the beam region, thereby generating a strong charge asymmetry of the muon halo.

\subsection{Detector}
\label{sec:detector}

The momenta of charged decay products were measured by a magnetic spectrometer, housed in a tank filled with helium at approximately atmospheric pressure, placed downstream of the decay volume. The spectrometer comprised four drift chambers (DCHs), each consisting of 8 planes of sense wires, and a dipole magnet located between the second and the third DCH which gave a horizontal transverse momentum kick of $265~\mathrm{MeV}/c$ to charged particles. The measured spectrometer momentum resolution was $\sigma_p/p = 0.48\%
\oplus 0.009\%\cdot p$, where the momentum $p$ is expressed in GeV/$c$. A counter hodoscope (HOD) consisting of two planes of orthogonal plastic scintillator strips producing fast charged particle trigger signals was placed after the spectrometer.

A 127~cm (27$X_0$) thick liquid krypton (LKr) electromagnetic calorimeter, used for lepton identification and as a photon veto detector in the present analysis, was located further downstream. Its 13248 readout cells had a transverse size of 2$\times$2 cm$^2$ each with no longitudinal segmentation. The energy resolution was $\sigma_E/E=3.2\%/\sqrt{E}\oplus9\%/E\oplus0.42\%$ ($E$ in GeV). The spatial resolution for the transverse coordinates $x$ and $y$ of an isolated electromagnetic shower was $\sigma_x=\sigma_y=0.42~{\rm cm}/\sqrt{E}\oplus0.06$~cm ($E$ in GeV).

A more detailed description of the detector components used for this measurement can be found in Ref.~\cite{fa07}.

\subsection{Trigger logic}
\label{sec:trigger}

A relatively low beam intensity (corresponding to $\sim10^5$ kaon decays in the vacuum tank per second) was used to enable the operation of a minimum-bias trigger configuration with high efficiency, and to minimize the accidental background. The $K_{e2}$ trigger condition consisted of the coincidence of signals in the two HOD planes (the $Q_1$ signal), loose lower and upper limits on the DCH hit multiplicity (the 1-track signal) and a LKr energy deposit of at least 10 GeV (the $E_{\rm LKr}$ signal). The $K_{\mu 2}$ trigger condition required a coincidence of the $Q_1$ and 1-track signals downscaled by a factor of $D=150$. The 1-track condition was not used in the early stage of the data taking (about 10\% of the total beam flux). Downscaled control triggers were collected to monitor the performance of the main trigger signals.

\section{Data samples}
\label{sec:samples}

The data used for this measurement were obtained from about $3.5\times 10^5$ SPS spills (with $\sim 10^{12}$ protons per spill), collected in 4 months of operation, and correspond to about $2\times10^{10}$ $K^\pm$ decays in the vacuum tank. The data-taking strategy was optimized to measure the two main backgrounds in the $K_{e2}$ sample, which are due to the beam halo muons and to $K_{\mu2}$ decays with a muon ($\mu^\pm$) misidentified as an electron ($e^\pm$).

As no kaon tracking is available, beam halo muons are a direct source of background to $K_{\mu2}$ decays, as well as to $K_{e2}$ decays via $\mu^\pm\to e^\pm\nu_e\nu_\mu$ decays in the fiducial region ($\nu_\ell$ is used to denote either a neutrino or an antineutrino here and below). The muon scrapers installed in the beam line were optimized for halo background suppression in the $K^+_{\ell2}$ data samples (as quantified below), making the $K^+_{\ell2}$ decays more favourable for the measurement. To measure the muon halo background directly from data, the $K^+$ and $K^-$ data samples were collected alternately by blocking the negative or the positive beam, respectively. Therefore, 65\%
(8\%) of the total 2007 beam flux corresponded to $K^+$ ($K^-$) decays collected in single-beam mode. In addition to being the signal samples (i.e. providing the $K_{\ell 2}$ data), these data sets are used as control samples to measure the muon halo background to the decays of opposite sign kaons (see Section~\ref{sec:halo}). The remaining 27\% of the 2007 beam flux corresponded to $K^\pm$ decays collected with simultaneous beams with the ratio of kaon fluxes of $\Phi(K^+)/\Phi(K^-)\approx 2$, and cannot be used for the halo background subtraction. An additional $K^-$ data sample collected in 2008, corresponding to about 4\% of the total 2007 beam flux, is used for halo subtraction in the $K^+$ sample, but not for the $R_K$ measurement.

To estimate the $K_{\mu2}$ background, the probability to misidentify a muon as an electron due to large energy deposition in the LKr calorimeter has been measured. This required the collection of a muon sample free from the typical $\sim\!10^{-4}$ electron contamination due to $\mu^\pm\to e^\pm\nu_e\nu_\mu$ decays in flight. To this end, 55\% of the kaon flux in 2007 was collected with a transverse horizontal lead (Pb) bar installed below the beam pipe between the two HOD planes, approximately 1.2~m in front of the LKr calorimeter. The bar was $9.2X_0$ thick in the beam direction (including an iron holder) and shadowed 11 rows of LKr cells (about 10\% of the total number of cells). For a 50~GeV electron traversing the Pb bar, the probability of depositing over 95\% of its initial energy in the LKr is $\sim\!5\times 10^{-5}$, as estimated with a simulation.

Due to the different acceptance and background conditions, $K^+$ and $K^-$ decays, as well as data collected with and without the Pb bar, are analyzed separately. The four resulting independent data samples are denoted as $K^+$(Pb), $K^+$(noPb), $K^-$(Pb) and $K^-$(noPb). The earlier analysis~\cite{la11} is based on the $K^+$(noPb) data set only, which contains 41\% of the reconstructed $K_{e2}$ candidates and has the lowest background contamination. The present analysis extends to the whole data sample and involves several
improvements on the estimation of backgrounds and systematic uncertainties.

\section{Data analysis}

\subsection{Analysis strategy}
\label{sec:strategy}

The analysis is based on counting the numbers of reconstructed $K_{e2}$ and $K_{\mu 2}$ candidates collected simultaneously; therefore it does not rely on an absolute kaon flux measurement. As a consequence, several systematic effects cancel to first order. Due to the dependence of the acceptance and background on the lepton momentum, the $R_K$ measurement is performed independently in 10 lepton momentum bins covering a range from 13 to 65~GeV/$c$ (the lowest momentum bin spans 7 GeV/$c$, the others are 5 GeV/$c$ wide). Since the $K^+/K^-$ and Pb/noPb samples are treated independently, the analysis is performed separately for 40 statistically independent subsamples with partially correlated systematic uncertainties. The ratio $R_K$ in each subsample is computed as
\begin{displaymath}
R_K = \frac{1}{D}\cdot \frac{N(K_{e2})-N_{\rm B}(K_{e2})}{N(K_{\mu
2}) - N_{\rm B}(K_{\mu 2})}\cdot \frac{A(K_{\mu 2})}{A(K_{e2})}
\cdot \frac{f_\mu\times\epsilon(K_{\mu 2})}
{f_e\times\epsilon(K_{e2})}\cdot\frac{1}{f_\mathrm{LKr}},
\end{displaymath}
where $N(K_{\ell 2})$ are the numbers of selected $K_{\ell 2}$ candidates $(\ell=e,\mu)$, $N_{\rm B}(K_{\ell 2})$ are the numbers of background events, $A(K_{\mu 2})/A(K_{e2})$ is the ratio of the geometrical acceptances (the acceptance correction), $f_\ell$ are the lepton identification efficiencies, $\epsilon(K_{\ell 2})$ are the trigger efficiencies, $f_\mathrm{LKr}$ is the global efficiency of the LKr readout (affecting the $K_{e2}$ reconstruction only), and $D=150$ is the $K_{\mu 2}$ trigger downscaling factor.

To evaluate the acceptance correction and the geometrical parts of the acceptances for most background processes entering the computation of $N_{\rm B}(K_{\ell 2})$, a detailed Monte Carlo (MC) simulation based on Geant3~\cite{geant3} is used that includes the time variation of data-taking conditions. The particle identification, trigger and readout efficiencies, as well as the muon halo background, are measured directly from data.
The determination of each term in the above expression is discussed below.

\subsection{Event reconstruction and selection}
\label{sec:selection}

Charged particle trajectories and momenta are reconstructed from hits and drift times in the spectrometer using a detailed magnetic field map. Fine calibrations of the spectrometer field integral and DCH alignment are performed by monitoring the mean reconstructed $K^\pm\to\pi^\pm\pi^+\pi^-$ invariant mass and the mean reconstructed missing mass in $K_{\mu 2}$ decays. Clusters of energy deposition in the LKr calorimeter are found by locating the maxima in the digitized pulses from individual cells. Shower energies are corrected for energy outside the cluster boundary, energy lost in isolated inactive cells (0.8\% of the total number) and cluster energy sharing. The energy response is calibrated with samples of electrons and positrons from $K^\pm\to\pi^0e^\pm\nu$ decays. Further details about the reconstruction procedures can be found in Ref.~\cite{fa07}.

\begin{figure}[p]
\begin{center}
\resizebox{0.5\textwidth}{!}{\includegraphics{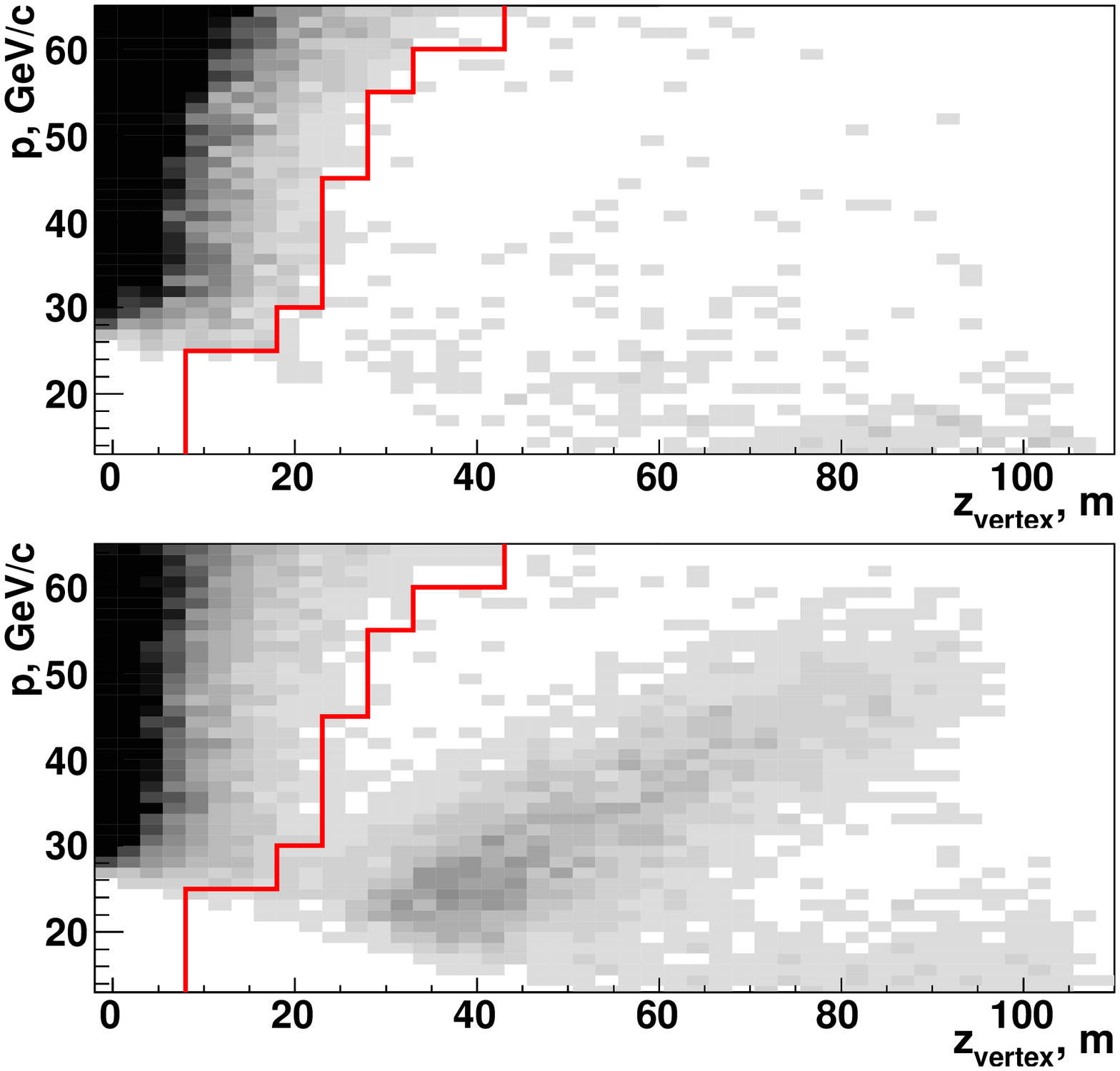}}%
\resizebox{0.5\textwidth}{!}{\includegraphics{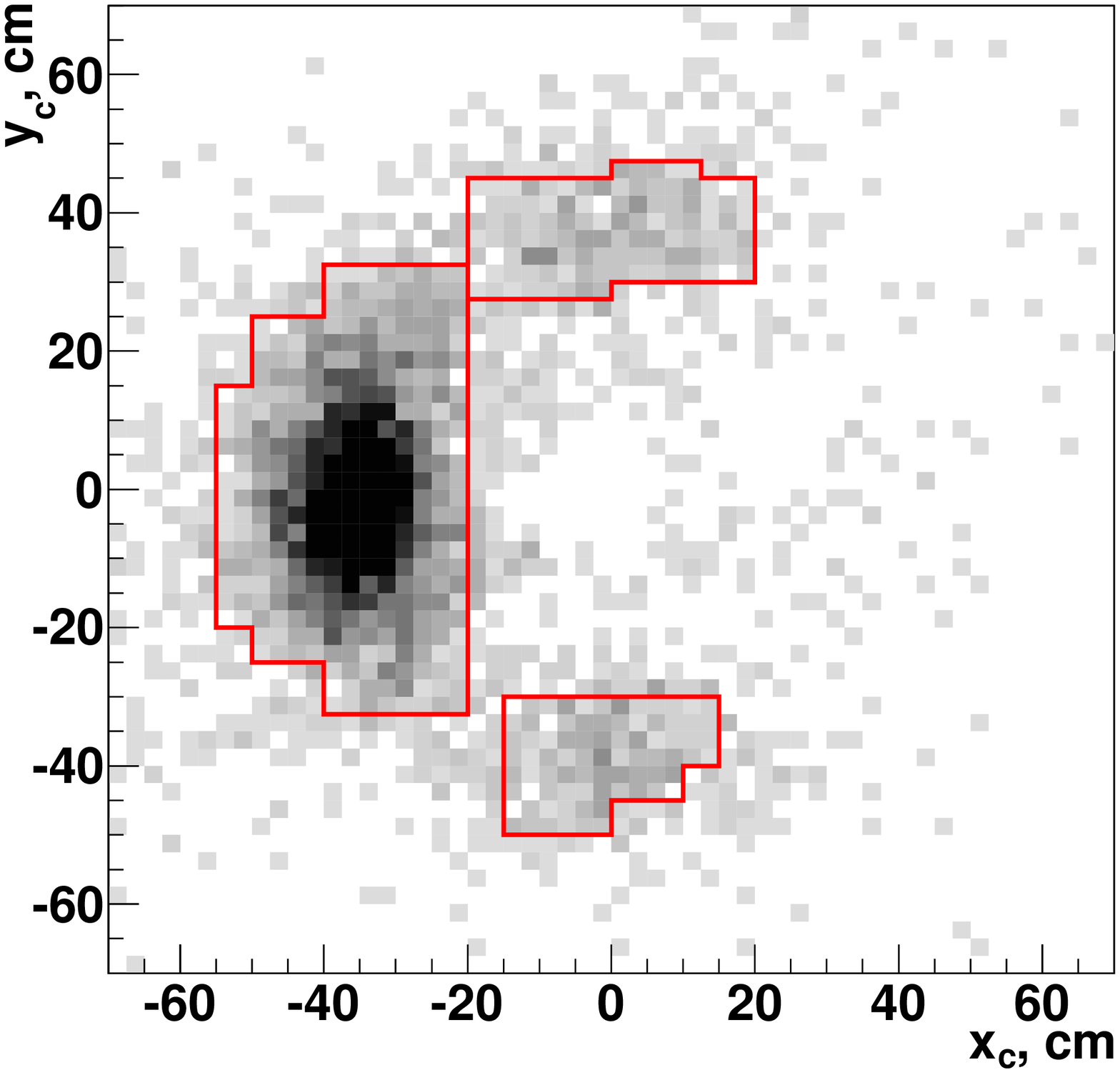}}%
\put(-250,201){\bf\large (a)} \put(-250,94){\bf\large (b)}
\put(-21,201){\bf\large (c)}
\end{center}
\vspace{-14mm} \caption{Properties of the muon halo background to $K^\pm_{e2}$(noPb) decays measured using control data samples as explained in Section~\ref{sec:halo}. Distributions of the reconstructed (a) $e^+$ and (b) $e^-$ candidate momentum versus the reconstructed longitudinal coordinate of the kaon decay vertex. The momentum-dependent $z_{\rm min}$ parameter optimized for muon halo suppression is indicated by solid lines: events to the left of the lines are rejected. The residual background in the $K^-_{e2}$ sample is a factor of 5 higher than in the $K^+_{e2}$ sample. (c) Distribution of the $e^-$ candidate crossing points ($x_c$, $y_c$) at the final collimator plane ($z=0$) for the halo background in the $K^-_{e2}$(noPb) sample surviving the $z_{\rm min}$ cut. The regions enclosed within solid lines are excluded to suppress the background. Halo background distributions for the $K^\pm_{e2}$(Pb) samples are similar.} \label{fig:halo}
\end{figure}

Most selection criteria are common to both the $K_{e2}$ and $K_{\mu2}$ decay modes. The principal criteria are the following.
\begin{itemize}
\item Exactly one reconstructed charged particle track (lepton candidate) geometrically consistent with originating from a kaon decay is required. The geometrical consistency is determined by reconstructing the decay vertex as the point of closest approach of the lepton candidate track extrapolated upstream (taking into account the measured stray magnetic field in the vacuum tank) and the axis of the kaon beam of the corresponding charge (determined with fully reconstructed $K^\pm\to\pi^\pm\pi^+\pi^-$ decays). The reconstructed closest distance of approach (CDA) of the lepton track to the beam axis is required to be less than 3.5~cm, which is crucial for the suppression of the muon halo background, as well as backgrounds from $K^\pm$ decays followed by pion or muon decays in flight. The CDA computation also defines the decay vertex.
\item The track impact points in the DCHs, HOD and LKr calorimeter must be within the corresponding fiducial acceptances, including appropriate separations from detector edges and inactive LKr cells. Moreover, it is impossible to efficiently identify electrons traversing the Pb bar by energy deposition in the LKr. Therefore, the region of the LKr calorimeter shadowed by the bar is excluded from the geometrical acceptance for the $K^\pm$(Pb) data samples, which leads to a reduction of the acceptance by $18\%$ for both $K_{e2}$ and $K_{\mu2}$ decays for these samples.
\item The reconstructed lepton momentum must be in the range 13 to 65~GeV/$c$. The lower limit ensures high efficiency of the $E_\mathrm{LKr}$ energy deposit trigger condition (see Section~\ref{sec:trigger}). Above the upper limit, the analysis is affected by larger systematic uncertainties due to backgrounds, as most backgrounds discussed in Section~\ref{sec:bkgs_all} increase at high track momentum.
\item No LKr clusters with energy above 2~GeV and within 12~ns of the track time are allowed, unless they can be associated to the track via direct energy deposition or bremsstrahlung. This requirement provides a photon veto to suppress backgrounds from $K^\pm\to e^\pm\nu\gamma$, $K^\pm\to\pi^0e^\pm\nu$ and $K^\pm\to\pi^\pm\pi^0$ decays. In the $K^\pm$(Pb) data samples, the LKr clusters located in the shadow of the bar are not used for the veto condition, as the LKr response for photons traversing the bar is difficult to be reproduced by simulation. This reduces the photon veto coverage and increases backgrounds from the above decays, as quantified in Section~\ref{sec:bkgs}.
\item The reconstructed kaon decay vertex must be located within the vacuum decay volume: its longitudinal coordinate $z_{\rm vertex}$ must satisfy the condition $z_{\rm vertex}>z_{\rm min}$. Here $z_{\rm min}$ depends on the reconstructed lepton momentum and is optimized for the suppression of the muon halo background, as described in Section~\ref{sec:halo} and illustrated for the $K^\pm_{e2}$(noPb) sample in Fig.~\ref{fig:halo}a,b. This requirement removes $\sim 95\%$ ($\sim 80\%$) of the halo background in $K^+_{e2}$ ($K^-_{e2}$) samples and decreases the geometrical acceptance for $K_{\ell 2}^\pm$ events by about 25\%.
\item The residual muon halo background affecting the $K^-_{\ell2}$ samples (Fig.~\ref{fig:halo}b) is found to have specific geometrical properties and can be strongly reduced by suitable cuts. The points ($x_c$, $y_c$) defined by extrapolation of the lepton track to the final collimator plane (a transverse plane located at the beginning of the vacuum decay volume, $z=0$) have a localized distribution, as shown for the $K^-_{e2}$(noPb) sample in Fig.~\ref{fig:halo}c. The optimization of a common selection condition for $K_{e2}^-$ and $K_{\mu2}^-$ decays (as required to minimize the bias on $R_K$) has been driven by the distribution of the background in the $K^-_{e2}$ sample, which is higher than the one in the $K^-_{\mu2}$ sample. The regions with highest population of background to the $K^-_{e2}$ decay indicated in Fig.~\ref{fig:halo}c are rejected, which reduces the halo background in the $K_{e2}^-$ samples by about 75\%, while decreasing the geometrical acceptance for $K_{\ell 2}^-$ events by about 10\%.
\end{itemize}

\begin{figure}[p]
\begin{center}
\resizebox{0.50\textwidth}{!}{\includegraphics{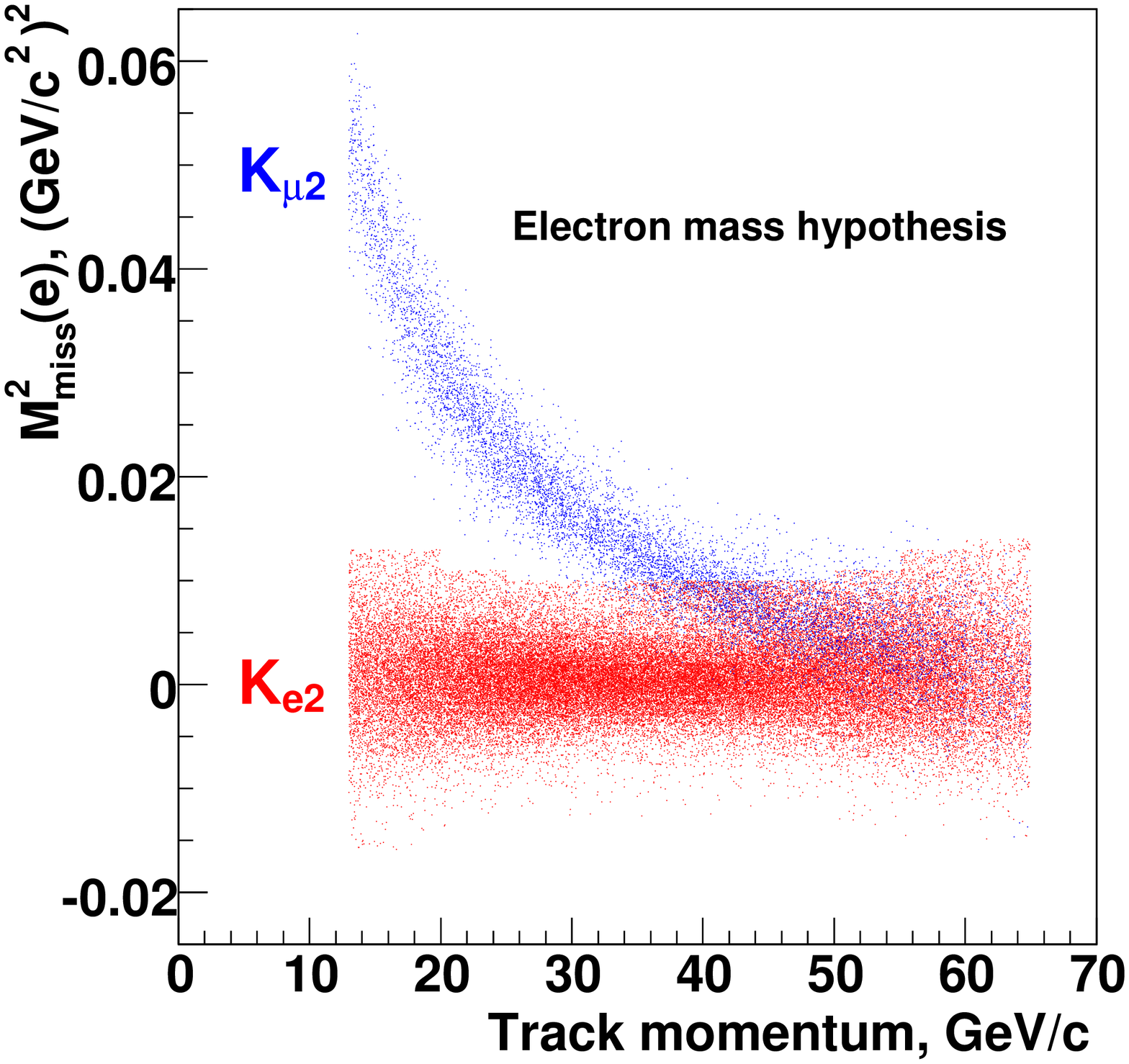}}%
\resizebox{0.50\textwidth}{!}{\includegraphics{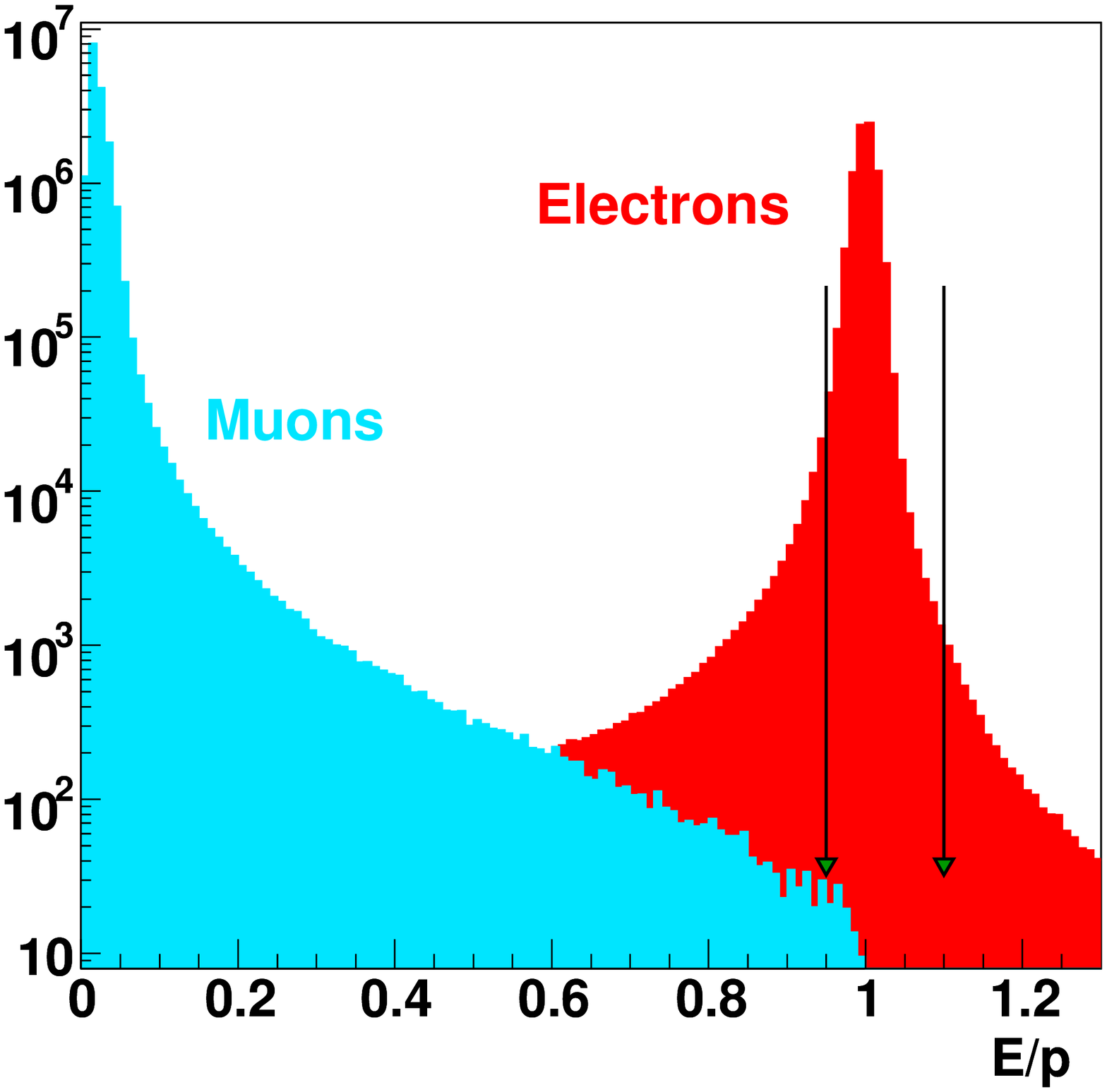}}
\put(-254,200){\bf\large (a)} \put(-27,200){\bf\large (b)}
\end{center}
\vspace{-14mm} \caption{(a) Reconstructed squared missing mass in the electron mass hypothesis $M^2_{\rm miss}(e)$ as a function of lepton momentum for $K_{e2}$ and $K_{\mu2}$ decays (data). The wrong mass assignment for the $K_{\mu2}$ decays leads to the momentum-dependence of $M^2_{\rm miss}(e)$. (b) $E/p$ spectra of electrons and muons (data) measured from $K^\pm\to\pi^0e^\pm\nu$ and $K_{\mu2}$ decays. The part of the muon spectrum above $E/p=0.95$ is for the muons traversing the Pb bar. The electron identification criterion applied for $p>25$~GeV/$c$ is indicated with arrows.} \label{fig:selection}
\end{figure}

Kinematic identification of the $K_{\ell 2}$ decays is based on the reconstructed squared missing mass assuming the track to be an electron or a muon: $M_{\mathrm{miss}}^2(\ell) = (P_K - P_\ell)^2$, where $P_K$ and $P_\ell$ ($\ell = e,\mu$) are the kaon and lepton 4-momenta, with $m_\ell$ used to assign the components of $P_\ell$. $P_K$ is taken from the average beam momenta, which are monitored with fully reconstructed $K^\pm\to\pi^\pm\pi^+\pi^-$ decays. Fig.~\ref{fig:selection}a shows the squared missing mass $M_{\rm miss}^2(e)$ evaluated in the electron hypothesis for $K_{e2}$ and $K_{\mu2}$ events as a function of lepton momentum. A selection condition $-M_1^2<M_{\mathrm{miss}}^2(\ell)<M_2^2$ is applied. The limits $M_1^2$ and $M_2^2$ vary across lepton momentum bins, taking into account the resolution on $M_{\mathrm{miss}}^2(\ell)$, radiative tails and background conditions. $M_1^2$ varies from 0.013~(GeV/$c^2$)$^2$ in the central region of the lepton momentum range to 0.016~(GeV/$c^2$)$^2$ at low and high momenta. Similarly, $M_2^2$ varies from 0.010 to 0.013~(GeV/$c^2$)$^2$ for $K^\pm$(noPb) samples and from 0.010 to 0.011~(GeV/$c^2$)$^2$ for $K^\pm$(Pb) samples. The latter limits are stricter to compensate for the weaker photon veto.

Lepton identification is based on the ratio $E/p$ of energy deposition in the LKr to momentum measured by the spectrometer, as illustrated in Fig.~\ref{fig:selection}b. Charged particles are identified as electrons if $(E/p)_\mathrm{min}<E/p<1.1$, where $(E/p)_\mathrm{min}=0.95$ for $p>25$~GeV/$c$ and $(E/p)_\mathrm{min}=0.9$ otherwise. The relaxed condition at low lepton momentum is possible because backgrounds in the $K_{e2}$ sample due to $\mu^\pm$ and $\pi^\pm$ misidentification are rejected kinematically in that momentum range (as seen for the $K_{\mu2}$ background in Fig.~\ref{fig:selection}a). For $(E/p)_\mathrm{min}=0.95$, this criterion leads to an electron identification efficiency $f_e>99\%$ and a probability of misidentifying a muon as an electron of $\sim 4\times10^{-6}$. Charged particles with $E/p<0.85$ are classified as muons; the corresponding muon identification inefficiency is negligible ($1-f_\mu\approx 3\times 10^{-5}$).


\subsection{Backgrounds}
\label{sec:bkgs_all}

\boldmath
\subsubsection{Muon halo background in the $K_{\ell2}$
samples} \unboldmath \label{sec:halo}

The rate and kinematical distribution of the muon-halo-induced backgrounds are qualitatively reproduced by a dedicated simulation. However, the precision of this simulation is limited by the uncertainties on the fringe fields of the beam line magnets. Therefore, these backgrounds have been measured directly by reconstructing the $K^+_{\ell 2}$ ($K^-_{\ell 2}$) decay candidates from control samples of positive (negative) tracks collected with the positive (negative) beam blocked, as described in Section~\ref{sec:samples}. The control samples used for background subtraction in the four independent $K_{\ell 2}$ signal samples are mutually exclusive, leading to independent statistical uncertainties. These samples are normalized to have the same numbers of muon events with $-0.3~(\mathrm{GeV}/c^2)^2<M_\mathrm{miss}^2(\mu)<-0.1~(\mathrm{GeV}/c^2)^2$ as the data (such events are not compatible with a kaon decay and can only arise from a halo muon).

As a cross-check, the backgrounds in the $K_{e2}$ samples have also been evaluated with a hybrid MC technique, using the measured spatial, angular and momentum distributions of the halo muons and simulating their decays ($\mu^\pm\to e^\pm\nu_e\nu_\mu$) assuming unpolarised muons, as the polarization is unknown. The results agree with those obtained from the direct measurements.

The non-zero probability of reconstructing a $K_{e2}$ candidate due to the decay of an opposite sign kaon enhances the halo background estimates for the $K_{e2}$ samples. The
effect is more pronounced for the $K^-$ samples because the ratio of the $K^+$ to $K^-$ beam fluxes is $\sim 2$. A $K^\pm$ decay must result in at least three charged daughter particles to produce an opposite sign particle. Contributions from $K^\pm\to\pi^0_D\ell^\pm\nu$, $K^\pm\to\pi^\pm\pi^0_D$ and $K^\pm\to\ell^\pm\nu e^+e^-$~\cite{bi93} decays (where $\ell=e,\mu$, and $\pi^0_D\to e^+e^-\gamma$ denotes the Dalitz decay) have been identified and subtracted using MC simulations. The corresponding correction to the final result is negligible: $\Delta R_K/R_K\sim 10^{-4}$.

In addition to being charge asymmetric (as explained in Section~\ref{sec:beam} and shown in Fig.~\ref{fig:halo}a, b), the muon halo background is left-right
asymmetric (as illustrated in Fig.~\ref{fig:halo}c for the $K^-_{e2}$ sample), and therefore depends on the polarity of the spectrometer magnetic field. To reduce the combined statistical uncertainty of the halo background estimates, almost equal samples of signal and control data were taken with each spectrometer polarity. The residual polarity imbalance was corrected by assigning weights to the control samples.

As control data have been collected mostly without the Pb bar, parts of these $K^{\pm}$(noPb) samples are used to estimate background in the $K^{\pm}$(Pb) signal samples by reducing the geometrical acceptance as described in Section~\ref{sec:selection}. This minimizes the overall statistical uncertainty. The background to signal ratio in the region covered by the Pb bar is lower than in the region outside the Pb bar due its geometrical localization. Therefore the backgrounds in the $K^{\pm}$(Pb) samples are higher than those in the $K^{\pm}$(noPb) samples.

The $K_{\ell 2}$ selection procedure and geometrical acceptance are time-dependent due to the variations of the beam geometry and the presence of temporarily masked groups of LKr cells. This implies a possible difference in geometrical acceptance between control samples (recorded with blocked beam) and signal samples (recorded with beam). The choice of control samples, selection and reconstruction procedures have been optimized to minimize these differences.

Uncertainties on the muon halo background estimates are due to the limited size of the control samples, the uncertainty of their normalization caused by decays of beam kaons and pions upstream of the decay volume, and the time dependence of the geometrical acceptance.


\boldmath \subsubsection{$K_{\mu2}$ background in the $K_{e2}$
sample} \unboldmath \label{sec:km2_bkg}

The $K_{\mu2}$ background results mainly from muon misidentification ($E/p>0.95$) due to \mbox{`catastrophic'} bremsstrahlung in or immediately in front of the LKr calorimeter. It has been addressed by a dedicated measurement of the misidentification probability $P_{\mu e}$ based on samples of muons from $K_{\mu2}$ decays traversing the Pb bar collected simultaneously with the $K_{\ell 2}^\pm$(Pb) data samples. These samples are statistically independent from the signal $K_{\mu 2}$ samples, as the Pb bar region is excluded from the geometrical acceptance, as discussed in Section~\ref{sec:selection}. As noted in Section~\ref{sec:samples}, electrons passing through the Pb bar have a probability of only $\sim5\times10^{-5}$ of being identified. Thus samples of misidentified muons traversing the Pb bar are electron-free for all practical purposes. However, the misidentification probability ($P_{\mu e}^{\rm Pb}$) for a muon traversing the Pb bar differs from the probability ($P_{\mu e}$) for an unintercepted muon because of ionization energy loss (dominant at low momentum) and bremsstrahlung (dominant at high momentum).

\begin{figure}[tb]
\begin{center}
\resizebox{0.50\textwidth}{!}{\includegraphics{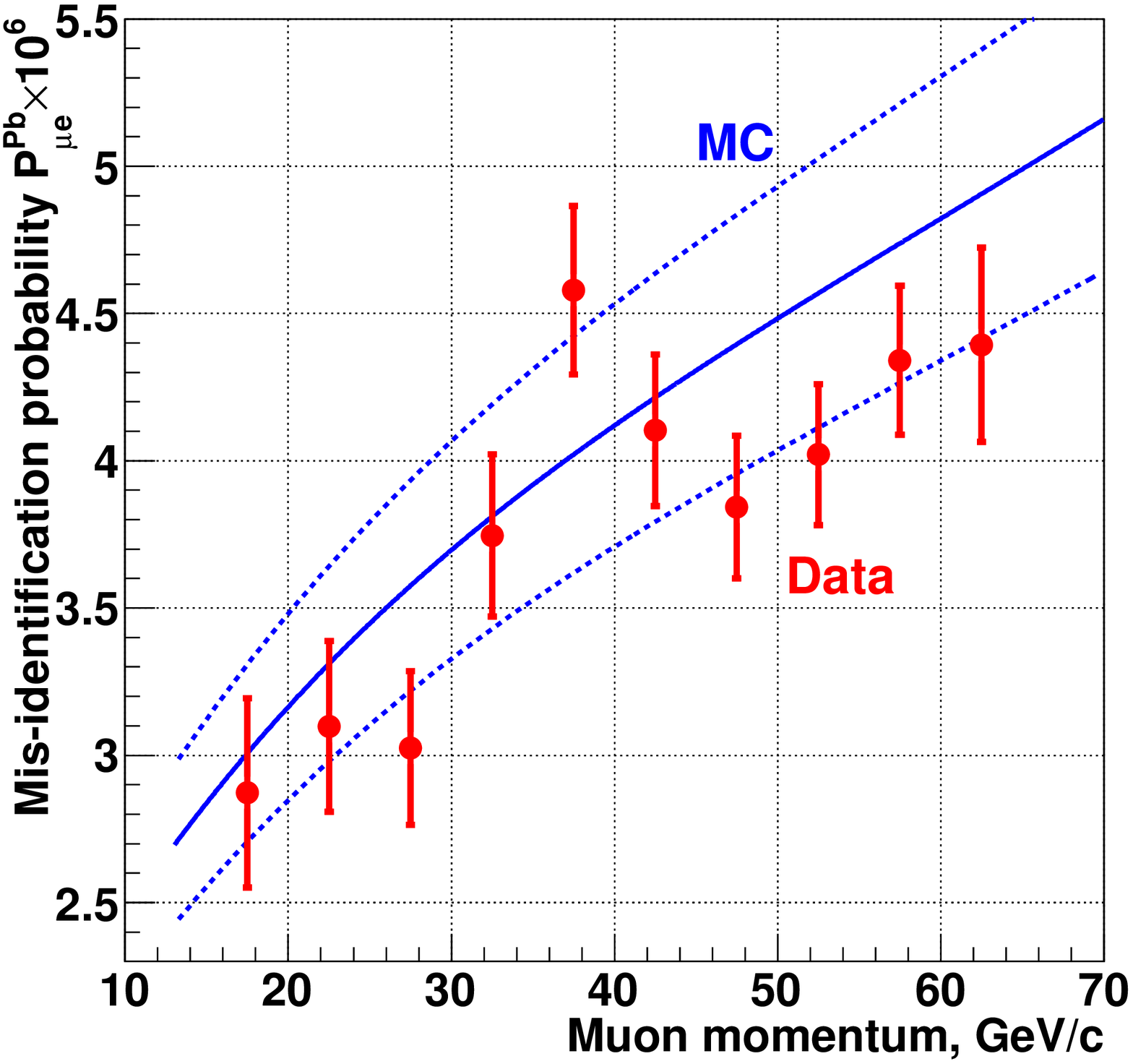}}%
\resizebox{0.50\textwidth}{!}{\includegraphics{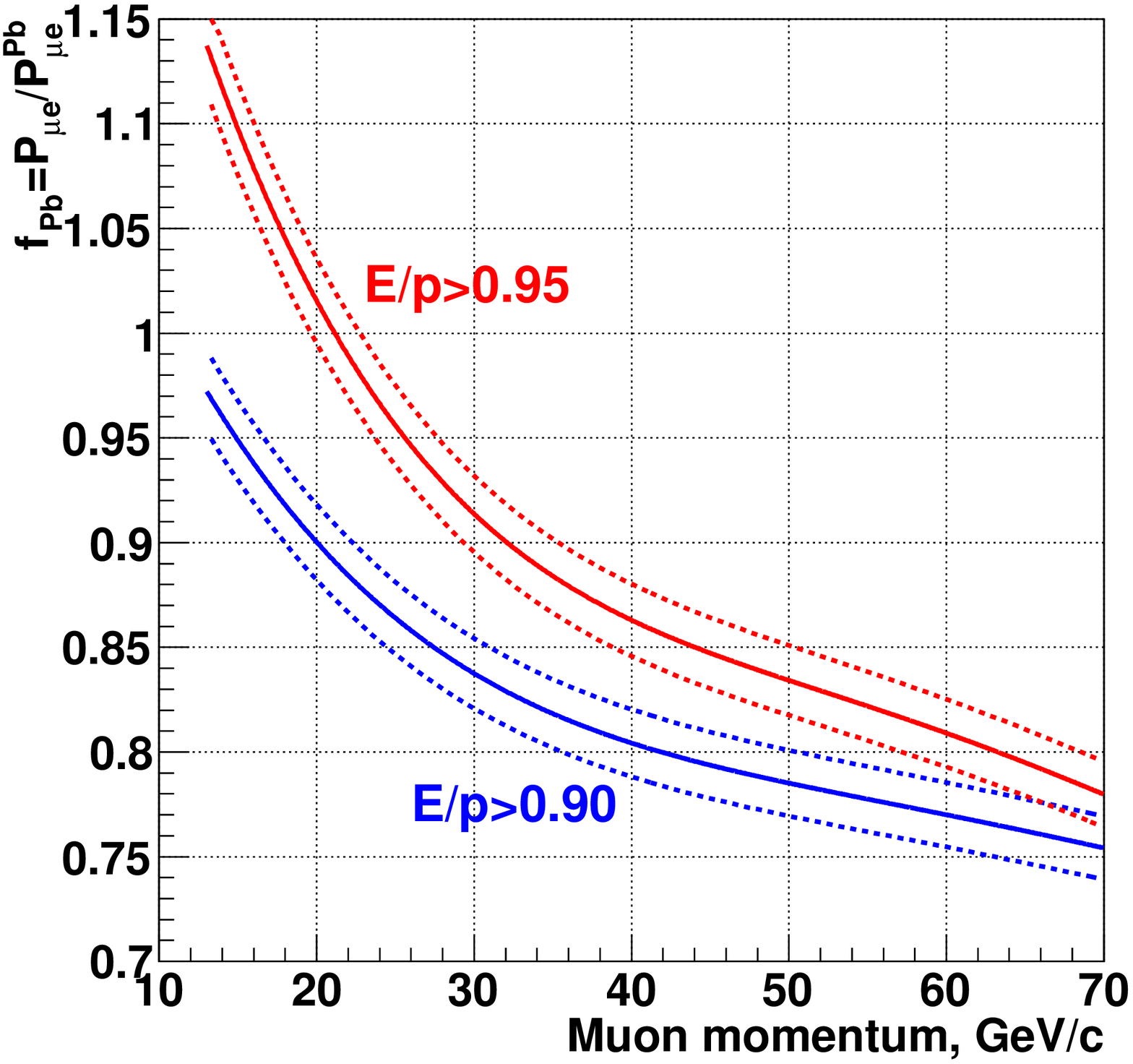}}
\put(-253,197){\bf\large (a)} \put(-26,197){\bf\large (b)}
\end{center}
\vspace{-14mm} \caption{(a) Misidentification probability for muons traversing the lead bar, $P_{\mu e}^\mathrm{Pb}$, for $(E/p)_\mathrm{min}=0.95$ as a function of momentum: measurement (solid circles with error bars; the uncertainties are uncorrelated) and simulation (solid line). (b) Correction factors $f_\mathrm{Pb}=P_{\mu e}/P_{\mu e}^\mathrm{Pb}$ evaluated by a simulation for the two specified values of $(E/p)_\mathrm{min}$. Dotted lines in both plots indicate the estimated systematic uncertainties arising from the simulation. The correlation of the latter uncertainties across the momentum values leads to a consistent shift of the MC band with respect to the measurements; the data/MC agreement validates the assigned systematic uncertainties.}
\label{fig:pmue}
\end{figure}

To evaluate the corresponding momentum-dependent correction factor $f_\mathrm{Pb}=P_{\mu e}/P_{\mu e}^\mathrm{Pb}$ to the measured probability $P_{\mu e}^\mathrm{Pb}$ for muons traversing the Pb bar, a dedicated MC simulation based on Geant4 (version 9.2)~\cite{geant4} has been developed to describe the propagation of muons downstream of the spectrometer involving all electromagnetic
processes including muon bremsstrahlung according to Ref.~\cite{ke97}. The LKr calorimeter
reconstruction has been optimized for showers initiated by electrons
and photons and starting near its front surface, whereas showers
initiated by muon bremsstrahlung start throughout the detector
volume. The corresponding \mbox{systematic} uncertainties on $P_{\mu e}$ and $P_{\mu e}^{\mathrm{Pb}}$ from the simulation due to energy calibration
and cluster reconstruction have been estimated to be $10\%$ of their values. However, the uncertainty on the ratio $f_\mathrm{Pb}=P_{\mu e}/P_{\mu e}^\mathrm{Pb}$ is significantly smaller ($\delta f_\mathrm{Pb}/f_\mathrm{Pb}=2\%$) due to a cancellation of the main systematic effects. Moreover, $f_\mathrm{Pb}$ has low sensitivity to the model of muon bremsstrahlung cross-section used. The measurements of $P_{\mu e}^{\mathrm{Pb}}$ in momentum bins compared to the results of the MC simulation and the correction factors $f_\mathrm{Pb}$ obtained by the simulation, along with the estimated systematic uncertainties of the simulated values, are shown in Fig.~\ref{fig:pmue}.

The $K_{\mu 2}$ background contamination due to muon misidentification has been computed using the geometrical acceptance evaluated by simulation, the measured $P_{\mu
e}^\mathrm{Pb}$ and the correction for the effect of the Pb bar, $f_\mathrm{Pb}$, evaluated from the simulation described above. The uncertainty on the background estimate comes from the limited size of the data sample used to measure $P_{\mu e}^\mathrm{Pb}$ and the uncertainty $\delta f_\mathrm{Pb}$ of the MC correction $f_\mathrm{Pb}$ for ionization and bremsstrahlung in the Pb bar. Moreover, the positive correlation between the reconstructed $M_\mathrm{miss}^2(e)$ and $E/p$, which are both computed using the reconstructed track momentum, leads to an apparent dependence of $P_{\mu e}$ on $M_\mathrm{miss}^2(e)$. The corresponding correction is based on the knowledge of the muon energy deposition spectrum in the vicinity of $E/p=1$; its model dependence leads to an
additional systematic uncertainty on the $K_{\mu2}$ background. The stability of the result with respect to variations in the lepton identification procedure was checked as discussed in Ref.~\cite{la11}.

The $K_{\mu2}$ decay also contributes to the background via $\mu^\pm\to e^\pm\nu_e\nu_\mu$ decays in flight. This background has been evaluated with a MC simulation. Energetic forward secondary electrons compatible with $K_{e2}$ kinematics and topology are strongly suppressed by muon polarisation effects and are further suppressed by radiative corrections to the muon decay~\cite{ar02}. The background from muon track association with accidental LKr clusters has been measured to be negligible.


\boldmath \subsubsection{$K^\pm\to e^\pm\nu\gamma$ background in the
$K_{e2}$ sample} \unboldmath

The definition of $R_K$ includes IB but excludes structure-dependent (SD) radiation~\cite{ci07}. The $\mathrm{SD}^+$ (positive photon helicity) component of the $K^\pm\to e^\pm\nu\gamma$ process peaks at high electron momentum in the
$K^\pm$ rest frame ($E^*_e\approx M_K/2$)~\cite{bi93}; it is
therefore kinematically similar to $K_{e2}$ decay. It can contribute to the background if the photon escapes the acceptance of the LKr calorimeter. The background has been estimated by a MC simulation based on the measured $K^\pm\to
e^\pm\nu\gamma~(\mathrm{SD}^+)$ differential decay rate in the
kinematic region $E^*_e>200$~MeV~\cite{am09}. The main uncertainty
on this estimate is due to the limited experimental precision on the
decay rate.

The $K^\pm\to e^\pm\nu\gamma~(\mathrm{SD}^-)$ decay with negative
photon helicity peaking at $E^*_e\approx M_K/4$ is kinematically
incompatible with $K_{e2}$, and the corresponding background is
negligible. Similarly, the background from interference terms
between the IB and SD processes is negligible.

\boldmath \subsubsection{Other backgrounds in the $K_{e2}$ sample}
\unboldmath

The $K^\pm\to\pi^0e^\pm\nu$ and $K^\pm\to\pi^\pm\pi^0$ decays produce a $K_{e2}$ signature in two cases: a) all $\pi^0$ decay products are undetected and, for the latter decay mode, the $\pi^\pm$ is misidentified as $e^\pm$; b) the only reconstructed particle is an electron ($e^\pm$) from a Dalitz decay $\pi^0_D\to e^+e^-\gamma$.

Due to the significant missing mass, these decays can only be kinematically compatible with $K_{e2}$ if the kaon is in the high-momentum tail of the beam distribution, or the detected kaon decay daughter particle suffers large multiple scattering. The systematic uncertainties on these minor backgrounds are due to the limited precision of the simulation of the non-Gaussian tails of multiple
scattering; they have been estimated as 50\% of the contributions
themselves.

The estimation of the $K^\pm\to\pi^\pm\pi^0$ background involves the pion misidentification probability, which has been measured as a function of momentum from samples of $K^\pm\to\pi^\pm\pi^0$ and $K^0_L\to\pi^\pm e^\mp\nu$ decays (the latter collected during a special run). In particular, at high lepton momentum where $K^\pm\to\pi^\pm\pi^0$ contributes, the selection criterion $0.95<E/p<1.1$ leads to a misidentification probability $P_{\pi e}=(0.41\pm0.02)\%$.

Data samples collected with simultaneous $K^+$ and $K^-$ beams, namely the $K^-$(Pb) sample and a part of the $K^+$(Pb) sample, are affected by backgrounds due to decays of opposite sign kaons, in a way similar to the muon halo control samples described in Section~\ref{sec:halo}. Contributions from the
$K^\pm\to\pi^0\ell^\pm\nu$ and $K^\pm\to\pi^\pm\pi^0$ decays with subsequent $\pi^0_D\to\gamma e^+e^-$ decays have been identified and subtracted using MC simulations.


\begin{figure}[tb]
\begin{center}
\resizebox{0.5\textwidth}{!}{\includegraphics{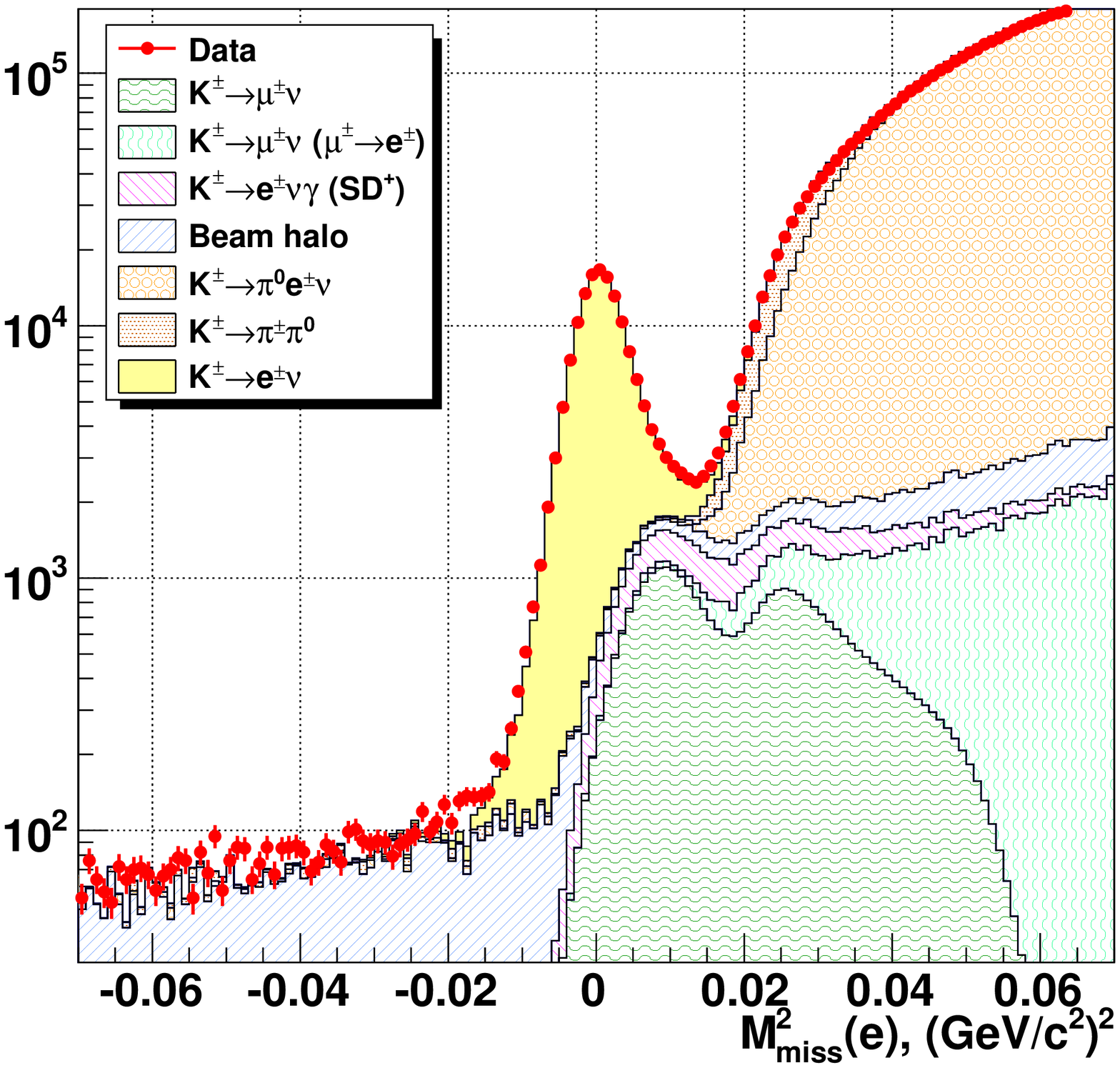}}%
\resizebox{0.5\textwidth}{!}{\includegraphics{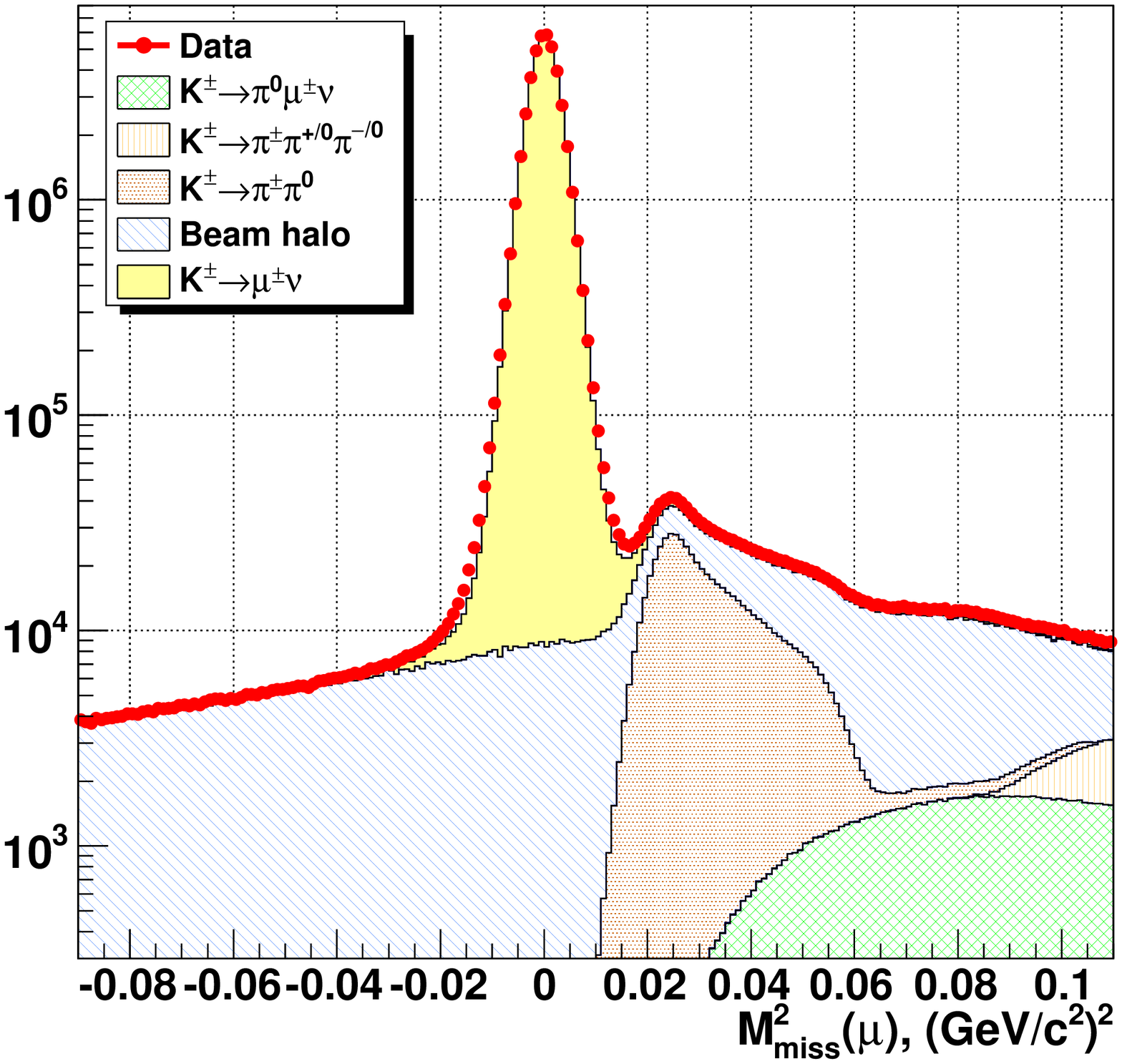}}%
\put(-248,201){\bf\large (a)} \put(-21,201){\bf\large (b)}
\end{center}
\vspace{-15mm} \caption{(colour online). Distributions of reconstructed squared missing masses (a) $M_{\mathrm{miss}}^2(e)$ and (b) $M_{\mathrm{miss}}^2(\mu)$ of the $K_{\ell 2}$ candidates compared with the sums of normalized estimated signal and background components. Beam halo contributions have been measured as discussed in Section~\ref{sec:halo}; the other contributions have been estimated with MC simulations involving the measured particle
misidentification probabilities. The double peak structure of the $K_{\mu2}$ background in the $K_{e2}$ sample originates from the momentum dependence of the electron identification condition.} \label{fig:mm2}
\end{figure}

\begin{table}[tb]
\begin{center}
\caption{Background contaminations in the $K_{\ell 2}$ samples
integrated over lepton momentum. The uncertainties on the background
in the $K_{\mu2}$ samples are negligible.} \vspace{1mm}
\label{tab:bkg}
\begin{tabular}{l|cccc}
\hline Data sample & $K^+$(noPb) & $K^+$(Pb) & $K^-$(noPb) & $K^-$(Pb)\\
\hline
$K_{e2}$ candidates & 59813 & 63282 & 10530 & 12333\\
Muon halo                                 & $(1.11\pm0.09)\%$ & $(1.51\pm0.10)\%$  & $(4.61\pm0.18)\%$  & $(7.86\pm0.23)\%$\\
$K_{\mu 2}$                               & $(6.11\pm0.22)\%$ & $(5.33\pm0.19)\%$  & $(5.76\pm0.20)\%$  & $(4.87\pm0.17)\%$\\
$K_{\mu 2}$ ($\mu\to e$ decay)            & $(0.26\pm0.04)\%$ & $(0.27\pm0.04)\%$  & $(0.31\pm0.09)\%$  & $(0.19\pm0.07)\%$\\
$K^\pm\to e^\pm\nu\gamma~(\mathrm{SD}^+)$ & $(1.07\pm0.05)\%$ & $(4.01\pm0.18)\%$  & $(1.25\pm0.06)\%$  & $(3.95\pm0.17)\%$\\
$K^\pm\to\pi^0 e^\pm\nu$                  & $(0.05\pm0.03)\%$ & $(0.28\pm0.14)\%$  & $(0.09\pm0.05)\%$  & $(0.37\pm0.17)\%$\\
$K^\pm\to\pi^\pm\pi^0$                    & $(0.05\pm0.03)\%$ & $(0.18\pm0.09)\%$  & $(0.06\pm0.03)\%$  & $(0.18\pm0.09)\%$\\
Opposite sign $K$                         & --                & $(0.04\pm0.01)\%$  & --                 & $(0.25\pm0.03)\%$\\
\hline
Total background                          & $(8.65\pm0.25)\%$ & $(11.62\pm0.33)\%$ & $(12.08\pm0.29)\%$ & $(17.67\pm0.39)\%$\\
\hline
$K_{\mu2}$ candidates $/10^6$& 18.027 & 18.433 & 3.069 & 3.288\\
Muon halo                    & 0.39\% & 0.44\% & 0.77\% & 1.22\%\\
\hline
\end{tabular}
\vspace{-10mm}
\end{center}
\end{table}


\subsubsection{Summary of backgrounds}
\label{sec:bkgs}

\begin{figure}[p]
\begin{center}
\resizebox{0.5\textwidth}{!}{\includegraphics{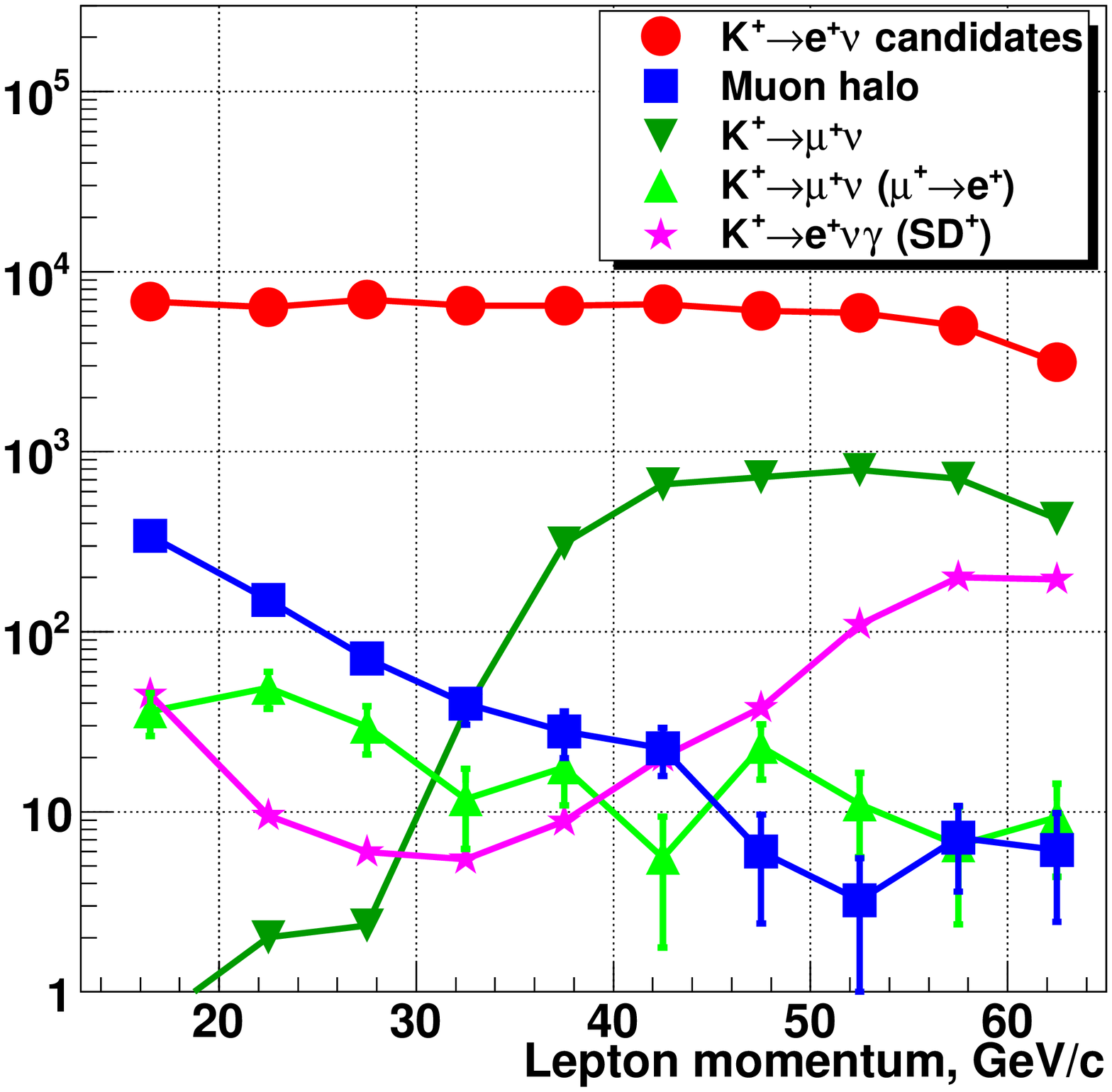}}%
\resizebox{0.5\textwidth}{!}{\includegraphics{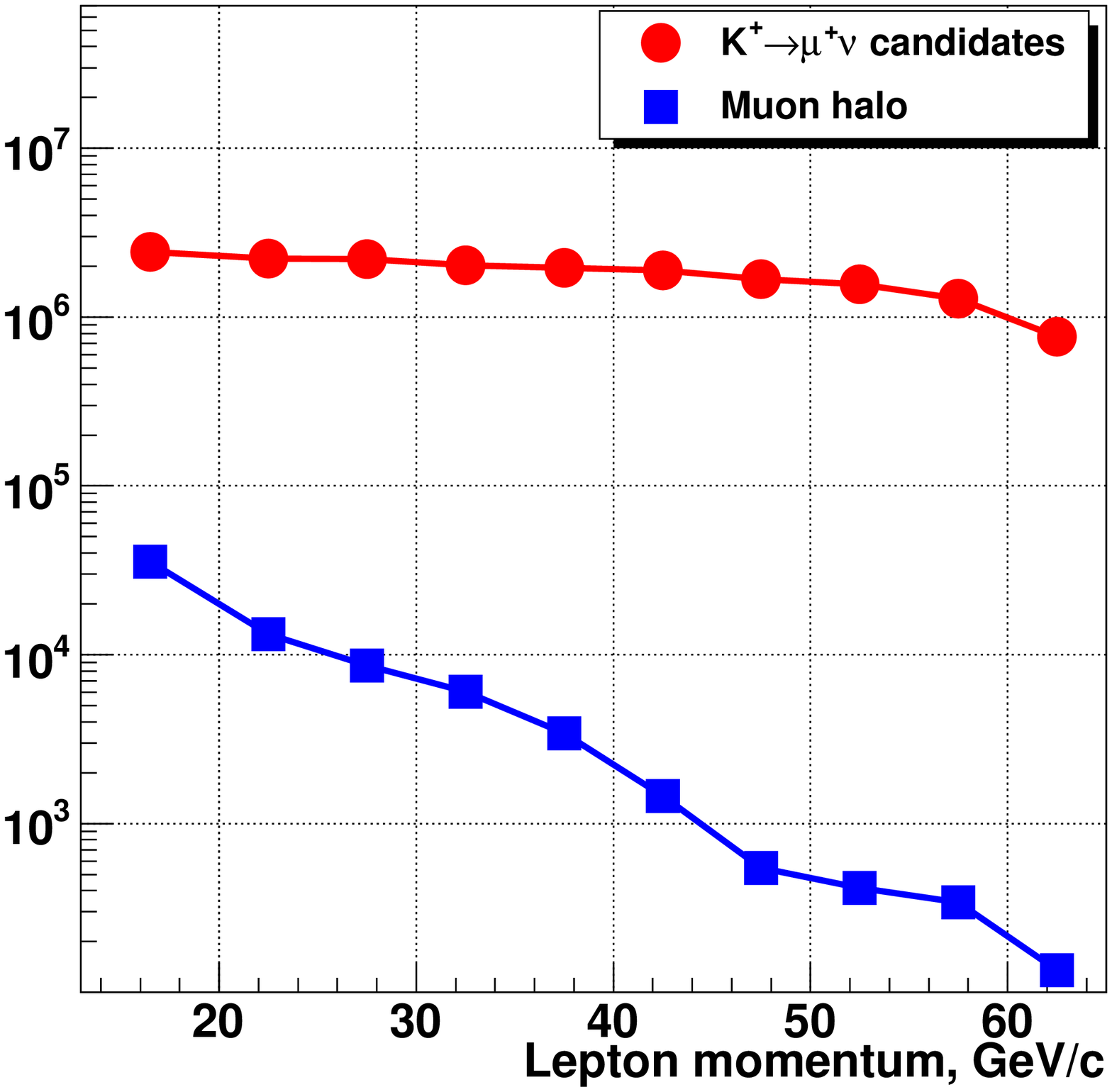}}%
\boldmath
\put(-429,202){\large $K^+_{e2}${\bf (noPb)}}%
\put(-202,202){\large $K^+_{\mu2}${\bf (noPb)}}\\
\unboldmath
\resizebox{0.5\textwidth}{!}{\includegraphics{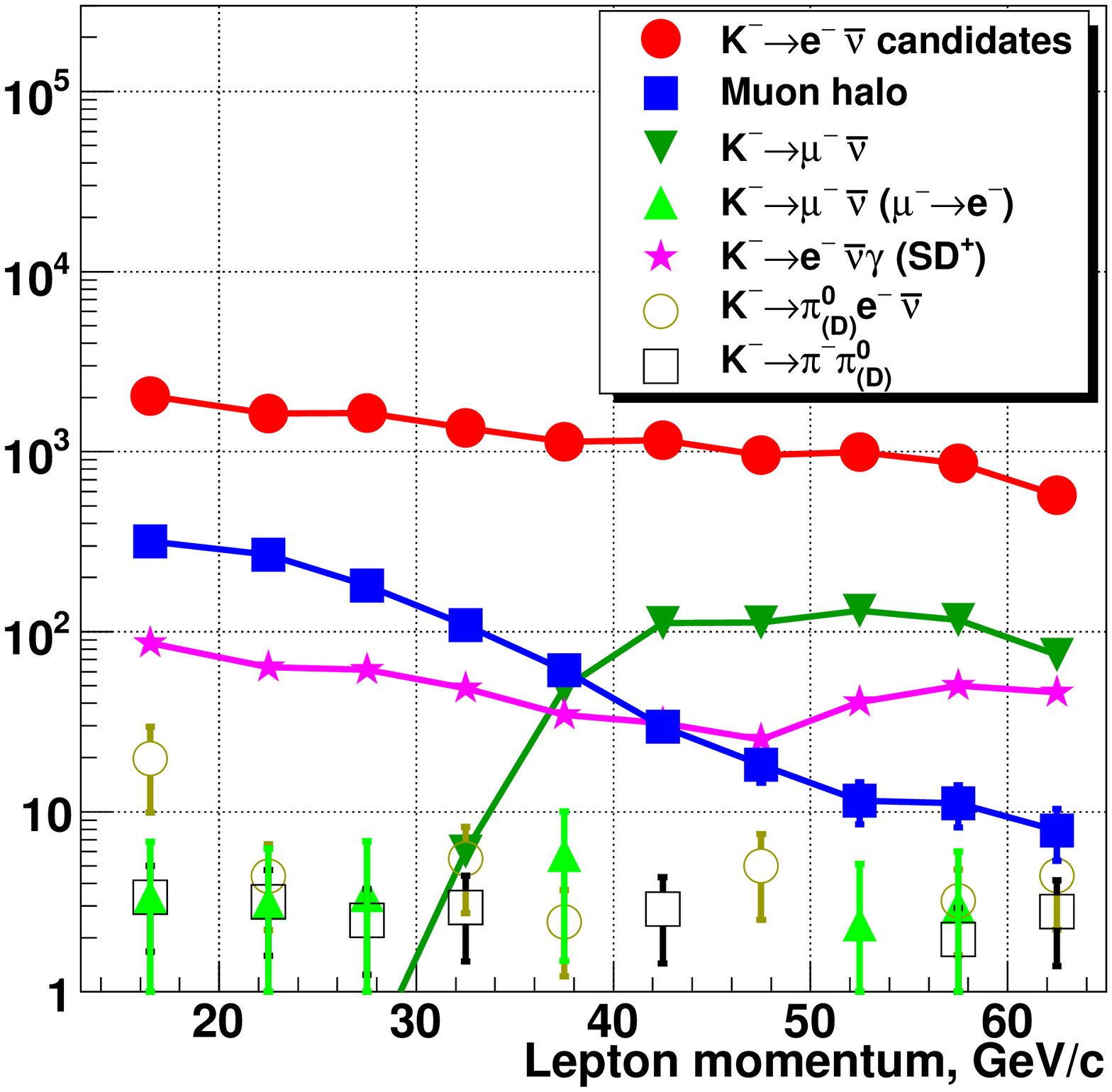}}%
\resizebox{0.5\textwidth}{!}{\includegraphics{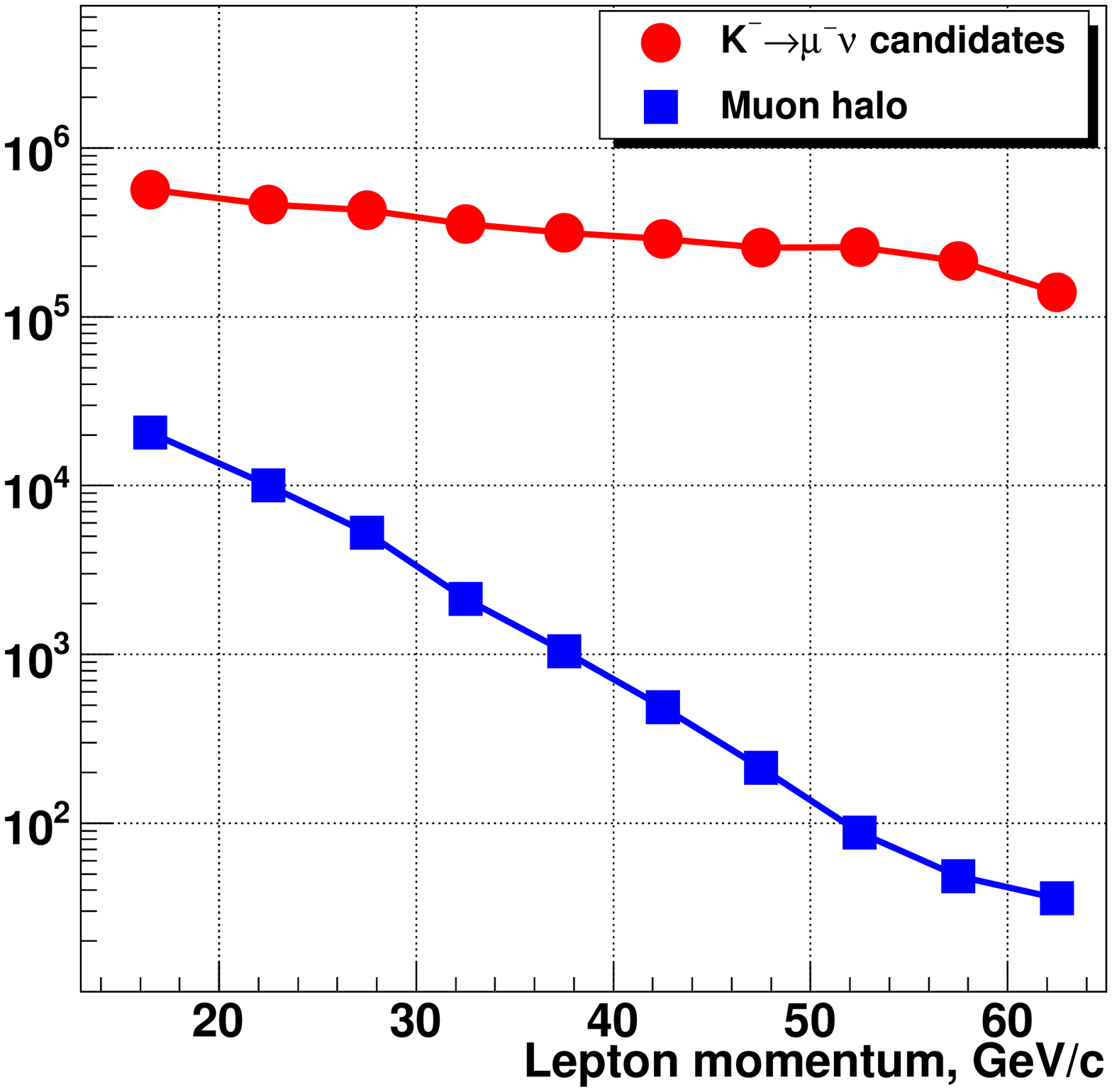}}%
\boldmath
\put(-429,202){\large $K^-_{e2}${\bf (Pb)}}%
\put(-202,202){\large $K^-_{\mu2}${\bf (Pb)}}%
\unboldmath
\end{center}
\vspace{-12mm} \caption{The numbers of $K_{\ell 2}$ candidates and main
background contributions in lepton momentum bins in the $K^+_{\ell 2}$(noPb) and
$K^-_{\ell 2}$(Pb) data samples, which have the lowest and the highest background contaminations, respectively. The differences between the data samples
in terms of the background composition are clearly visible.} \label{fig:bkgs}
\end{figure}

The $M_\mathrm{miss}^2(\ell)$ spectra of the selected $K_{\ell 2}$ candidates summed over all data samples are presented in Fig.~\ref{fig:mm2}. The numbers of the selected $K_{\ell 2}$ candidates and backgrounds in data samples integrated over the lepton momentum are summarized in Table~\ref{tab:bkg}. The total $K_{e2}$ sample consists of 145958 candidates with an estimated background of $(10.95\pm0.27)\%$. The dependences of the backgrounds on the lepton momentum for the $K_{\ell 2}^+$(noPb) and $K_{\ell 2}^-$(Pb) data samples, which have the lowest and the highest background contaminations, respectively, are displayed in
Fig.~\ref{fig:bkgs}. The total $K_{\mu 2}$ sample collected with a
pre-scaled trigger consists of $4.282\times 10^7$ candidates with a
background due to the beam halo muons measured to be $(0.50\pm0.01)\%$.

\subsection{Other systematic effects}
\label{sec:syst}

A detailed discussion of the systematic effects not related to
background subtraction is available in~\cite{la11}. The main points
are summarized below.


\subsubsection{Acceptance correction}

The typical geometric acceptance of the adopted $K_{\mu2}$ selection is about 34\% (42\%) for the data samples collected with (without) the Pb bar. The acceptance correction $A(K_{\mu2})/A(K_{e2})$ for each subsample, varying from 1.15 to 1.42 over the subsamples, has been evaluated by the MC simulation described in Section~\ref{sec:strategy}. Part of the correction, particularly at low lepton momentum, arises because of the wide radial distribution of electrons from $K_{e2}$ decays (not fully contained within the geometric acceptance of the DCH planes). Radiative effects result in a further loss of $K_{e2}$ acceptance by increasing the reconstructed $M_{\rm miss}^2(e)$ as quantified below. The radiative process $K^\pm\to e^\pm\nu\gamma$ from IB, which decreases the $K_{e2}$ acceptance by about 5\% relative, is included in the simulation following~\cite{bi93,ga06}. The evaluation of the correction for the external bremsstrahlung, which accounts for about 6.5\% relative $K_{e2}$ acceptance loss, requires a precise description of the material upstream of the spectrometer magnet~\cite{fa07}: a $\rm{Kevlar}\textsuperscript{\textregistered}$ window separating the vacuum decay volume from the spectrometer ($0.3\%X_0$), helium in the 12.3~m long volume ($0.4\%X_0$) and the two DCHs ($0.8\%X_0$).

The validity of the material description has been demonstrated by measuring the brems\-strah\-lung rate for a sample of $K^\pm\to\pi^0 e^\pm\nu(\gamma)$ decays collected concurrently with the main data set. The fraction of events in this sample with a reconstructed bremsstrahlung photon of at least 10 GeV energy, $f_\gamma$, has been used to measure the material thickness. The quantity $f_\gamma$ ranges from 0.3\% to 2\% for the $e^\pm$ momentum ranging from 20 GeV/$c$ to 40 GeV/$c$, with contributions from external bremsstrahlung ($\sim 70\%$) and $K^\pm\to\pi^0 e^\pm\nu\gamma$ (IB) decays ($\sim 30\%$). The relative uncertainty on the material thickness simulation has been conservatively estimated to be 1.1\% by comparing the values of $f_{\gamma}$ measured from the data and from simulated $K^\pm\to\pi^0 e^\pm\nu(\gamma)$ samples. The quoted uncertainty includes contributions from the limited size of the control sample and the precision of the $K^\pm\to\pi^0 e^\pm\nu\gamma$ (IB) decay simulation according to~\cite{ga06}. It translates into a systematic uncertainty of $\delta R_K=0.002\times 10^{-5}$ on the final result. No statistically significant variation of the material thickness over the data taking period (e.g. due to the helium purity variation) has been observed.


An independent validation of the material simulation has been performed using special data samples taken on two earlier occasions with low-intensity, mono-energetic $e^\pm$ beams steered into the spectrometer. The rates of the radiated photons (produced only by external bremsstrahlung) agree between data and simulation within 1\% precision.

Other sources of systematic uncertainty are the limited knowledge of the beam spectrum, profile and divergence, and the simulation of the LKr response to soft radiative photons. A separate uncertainty has been assigned to account for the finite precision of the spectrometer alignment.

\subsubsection{Electron identification efficiency}
\label{sec:electron_id}

Samples of electrons and positrons selected kinematically from $K^\pm\to\pi^0 e^\pm\nu$ and $K^0_L \to \pi^\pm e^\mp \nu$ decays have been used to calibrate the energy response of each LKr cell for each of the nine data taking periods defined for this purpose, as well as to measure the electron identification efficiency $f_e$ as a function of LKr cell, track momentum and time. The overall inefficiency averaged over the $K_{e2}$ sample is $1-f_e=(0.72\pm0.05)\%$, where the quoted uncertainty is mainly systematic. The corresponding correction to the final result is $\Delta R_K=(+0.018\pm0.001)\times 10^{-5}$.

The inefficiency has been found to be highly stable in time: averaged over the momentum ranges $p<25$ GeV/$c$ ($p>25$ GeV/$c$), it varies during the data taking from 0.43\% to 0.44\% (from 0.79\% to 0.81\%). Note that the electron identification criteria are different in the two momentum ranges, as explained in the last paragraph of Section~\ref{sec:selection}.


\subsubsection{Trigger and readout efficiencies}

The $(1.4\pm0.1)\%$ inefficiency of the $Q_1$ trigger condition, as well as the much smaller inefficiency of the 1-track trigger condition, nearly cancel between the $K_{e2}$ and $K_{\mu2}$ samples due to their geometrical uniformity and long-term stability. The
residual systematic bias is negligible. The inefficiency of the $E_{\rm LKr}$ condition, which enters only into the $K_{e2}$ trigger chain, has been measured with a sample of $K^\pm\to\pi^0 e^\pm\nu$ events. It is found to be significant only in the lowest bin of lepton momentum (close to the 10~GeV trigger energy threshold). It varies in that bin over time from 0.15\% to 0.5\%, and is localized geometrically. The correction to the final result for the $E_{\rm LKr}$ trigger inefficiency amounts to $\Delta R_K=+0.001\times 10^{-5}$, with a negligible uncertainty.

Energetic photons not reconstructed in the LKr may initiate showers by interacting with the DCHs or the beam pipe material, causing the DCH hit multiplicities to exceed the limits allowed by the 1-track trigger condition. The dominant background with a lost photon is from the $K^\pm\to e^\pm\nu\gamma~(\mathrm{SD}^+)$ decay. The corresponding 1-track trigger inefficiency for the background $K^\pm\to e^\pm\nu\gamma~(\mathrm{SD}^+)$ events in the $K_{e2}$ sample has been evaluated by a MC simulation. The simulation has been validated by comparison with a data sample of $K^\pm\to\pi^0 e^\pm\nu$ events with two lost photons. The inefficiency has been found to vary from 0.06 to 0.21 depending on the electron momentum, and its relative systematic uncertainty has been estimated to be 30\%. The correction to $R_K$ for this effect is $\Delta R_K=(+0.003\pm0.001)\times 10^{-5}$.

The misalignment of the LKr signal timing leads to a small global LKr readout inefficiency affecting the $K_{e2}$ reconstruction only. This inefficiency has been measured to be $1-f_\mathrm{LKr}=(0.20\pm0.03)\%$ and found to be stable in time, using an independent  readout system with larger granularity developed to monitor the trigger chain of the NA48 experiment~\cite{fa07}. The corresponding correction to the result is $\Delta R_K=(+0.005\pm0.001)\times 10^{-5}$.

\section{Averaging over the data samples}

\begin{figure}[tb]
\begin{center}
\resizebox{0.6\textwidth}{!}{\includegraphics{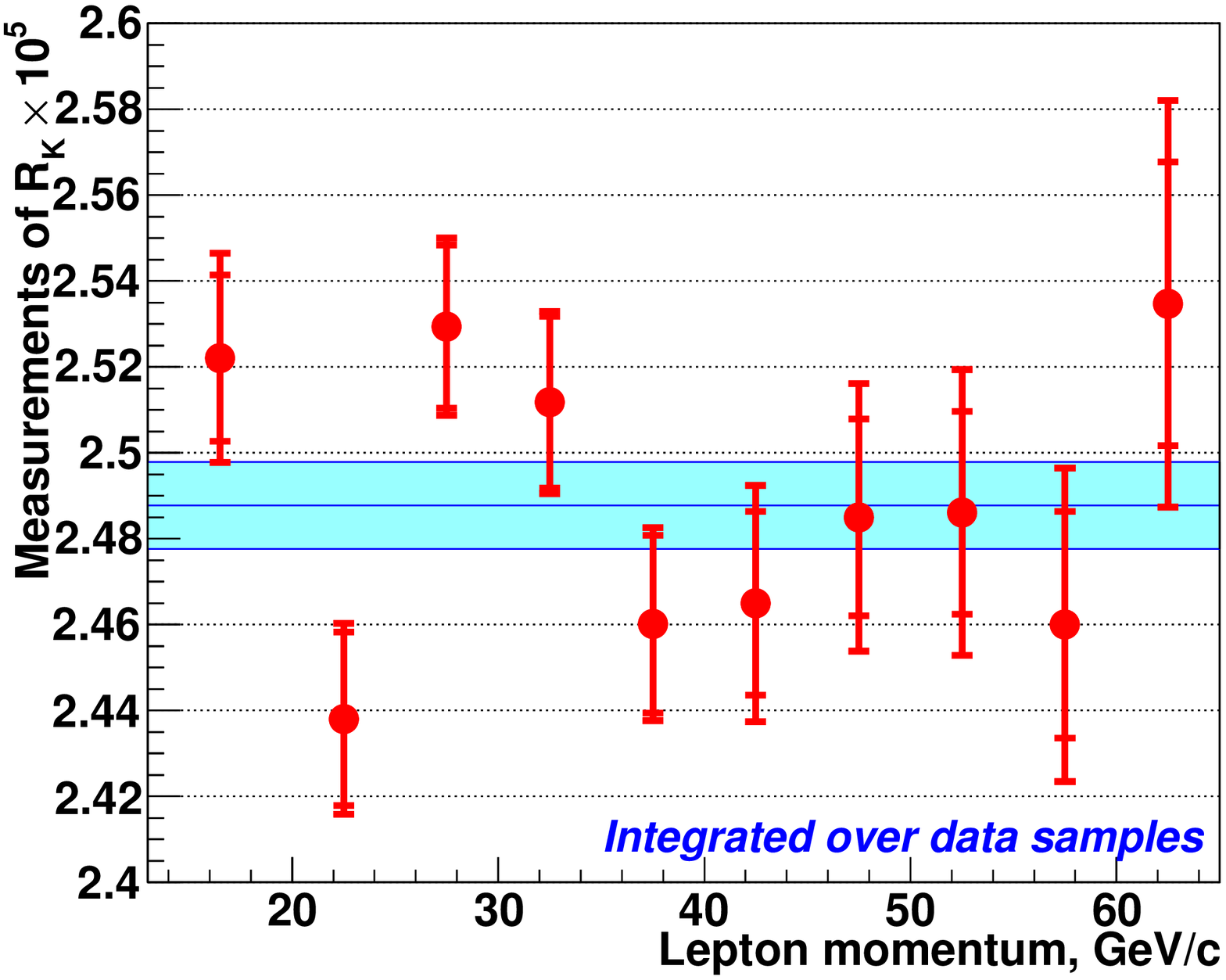}}%
\resizebox{0.4\textwidth}{!}{\includegraphics{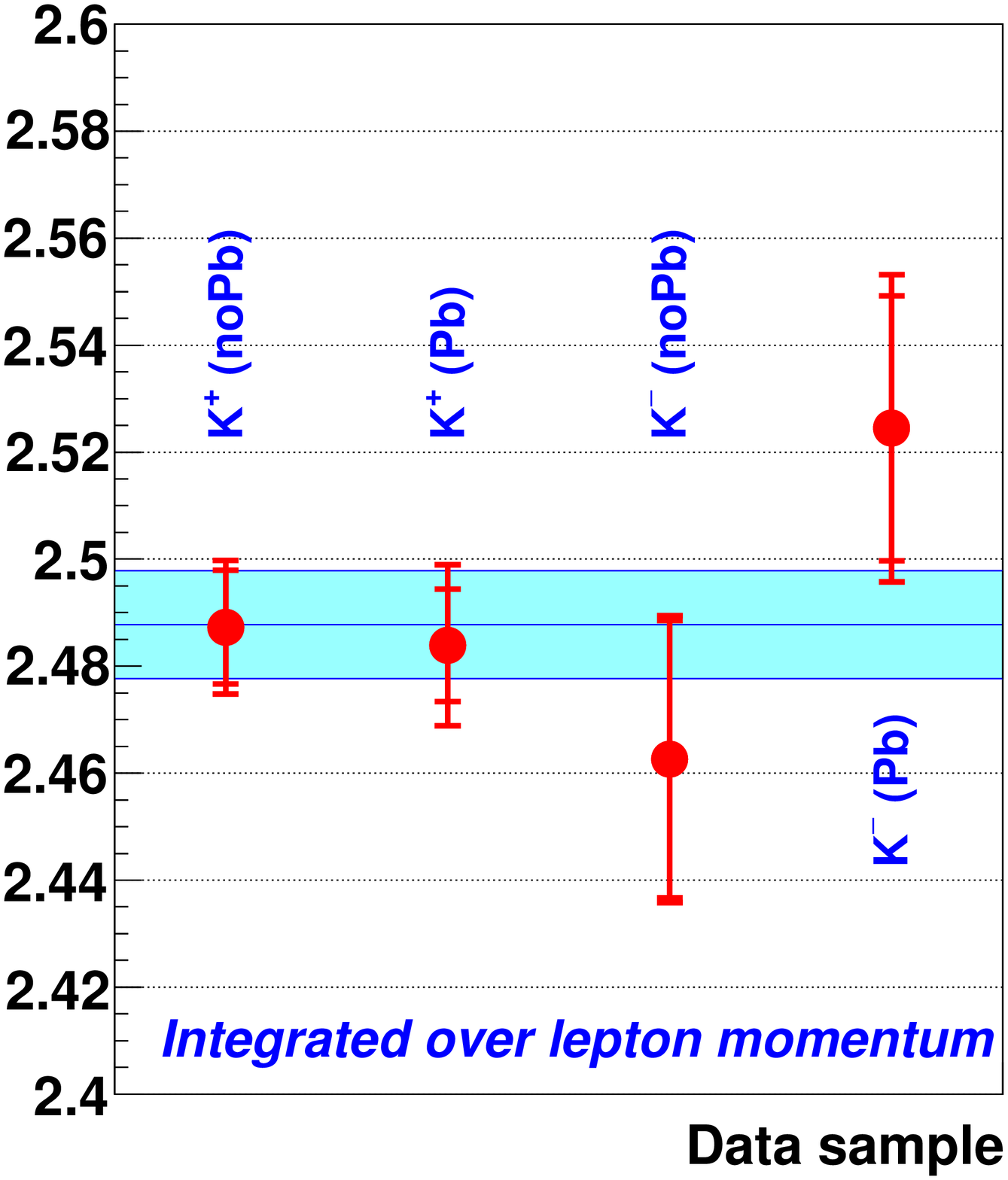}}%
\end{center}
\vspace{-14mm} \caption{Stability of the $R_K$ measurement versus
lepton momentum and for independent data samples. Statistical and
total errors of individual points are indicated with horizontal
dashes. The result of the $\chi^2$ fit to the 40 subsamples,
with the combined statistical and systematic error, is shown by the horizontal lines and bands.}\label{fig:rkfit}
\end{figure}

\begin{table}[tb]
\begin{center}
\caption{Summary of the uncertainties on $R_K$.}
\label{tab:err}\vspace{2mm}
\begin{tabular}{lc}
\hline
Source & $\delta R_K\times 10^5$\\
\hline
Statistical                                 & 0.007  \\
\hline
~~~$K_{\mu2}$ background                    & 0.004  \\
~~~$K^\pm\to e^\pm\nu\gamma~(\textrm{SD}^+)$ background         & 0.002\\
~~~$K^\pm\to\pi^0 e^\pm\nu$, $K^\pm\to\pi^\pm\pi^0$ backgrounds & 0.003\\
~~~Muon halo background                     & 0.002\\
~~~Spectrometer material composition        & 0.002\\
~~~Acceptance correction                    & 0.002\\
~~~Spectrometer alignment                   & 0.001\\
~~~Electron identification inefficiency     & 0.001\\
~~~1-track trigger inefficiency             & 0.001\\
~~~LKr readout inefficiency                 & 0.001\\
Total systematic                            & 0.007\\
\hline
Total & 0.010\\
\hline
\end{tabular}
\vspace{-1cm}
\end{center}
\end{table}

A $\chi^2$ fit to the 40 measurements of $R_K$ in individual
subsamples (consisting of the 4 data samples, each subdivided into
10 lepton momentum bins) has been performed. The following
correlations of the systematic uncertainties over subsamples have
been taken into account.
\begin{itemize}
\item All systematic uncertainties related to the acceptance correction and electron identification efficiency are conservatively considered to be fully correlated among data samples and lepton momentum bins.
\item The systematic uncertainties on the decay rates of all background kaon decay modes, coming from other measurements~\cite{pdg}, are fully correlated among data samples and lepton momentum bins. The most significant of these uncertainties is that of the $K^\pm\to e^\pm\nu\gamma~(\mathrm{SD}^+)$ decay rate.
\item The systematic uncertainties on the muon and pion misidentification probabilities have components which are fully correlated among the data samples (the statistical errors of independent $P_{\mu e}$ and $P_{\pi e}$ measurements in a given lepton momentum bin) as well as components which are fully correlated among both data samples and lepton momentum bins (due to the model-dependent correction described in Section~\ref{sec:km2_bkg}).
\item The systematic uncertainties on the muon halo background normalization, which is relevant only for high lepton momentum bins, are fully correlated among the data taking periods. Uncertainties due to the residual differences of geometrical acceptances among data and muon halo control samples are correlated among lepton momentum bins.
\item The systematic uncertainties due to the 1-track trigger inefficiency and LKr readout inefficiency are correlated among the data samples.
\item The systematic uncertainties on the $K^\pm\to\pi^0 e^\pm\nu$ and $K^\pm\to\pi^\pm\pi^0$ backgrounds due to imperfect simulation of multiple scattering at large angles and of the kaon beam spectrum, which affect the highest lepton momentum bin only, are correlated among the data samples.
\end{itemize}
The resulting correlation coefficients among the total uncertainties for the 40 subsamples range from 0.002 to 0.40. The fit result is
\begin{displaymath}
R_K = (2.488\pm 0.007_{\mathrm{stat.}}\pm
0.007_{\mathrm{syst.}})\times 10^{-5} =(2.488\pm0.010)\times
10^{-5},
\end{displaymath}
with $\chi^2/{\rm ndf}=47/39$ (probability 18\%). The stability of the measurements for bins of different lepton momentum (averaged over the four data samples) and for the different data samples (averaged over the ten momentum bins) is shown in
Fig.~\ref{fig:rkfit}. The contributions to the uncertainty of the result are
summarised in Table~\ref{tab:err}.


\section*{Summary}

The most precise measurement of $R_K=\Gamma(K_{e2})/\Gamma(K_{\mu 2})$ to date has been performed from a sample of 145958 $K_{e2}$ candidates with an estimated background of $(10.95\pm0.27)\%$ collected by the NA62 experiment in 2007--2008. The result $R_K=(2.488\pm0.010)\times10^{-5}$ is consistent with the earlier measurements and with the SM expectation. The experimental uncertainty on $R_K$ is still an order of magnitude larger than the uncertainty on the SM prediction, which motivates further measurements at improved precision.


\section*{Acknowledgements}

It is a pleasure to express our appreciation to the staff of the CERN laboratory and the technical staff of the participating laboratories and universities for their efforts in the operation of the SPS accelerator, the experiment and data processing. We gratefully acknowledge the excellent performance of the BlueBEAR computing facility at the
University of Birmingham, where the simulations have been performed. We thank H. Wahl for helpful discussion.



\end{document}